\documentclass[notitlepage,
 aip,
 amsmath,amssymb,
reprint,%
]{revtex4-1}

\usepackage[version=3]{mhchem} 
\usepackage{multirow}
\usepackage{threeparttable}
\usepackage[labelsep=period]{caption}
\usepackage{appendix}
\usepackage{braket}
\usepackage{caption}
\usepackage{subcaption}
\usepackage{comment}
\usepackage{color}
\usepackage{graphicx}
\usepackage{dcolumn}
\usepackage{bm}
\usepackage[section]{placeins}
\usepackage{soul}
\usepackage{siunitx}
\usepackage[table]{xcolor}
\usepackage{amsmath}

\usepackage[labelsep=period]{caption}
\renewcommand{\thetable}{\arabic{table}}
\usepackage{multirow}
\usepackage{threeparttable}
\usepackage{eucal}

\begin{document}

\preprint{AIP/123-QED}

\title[]{
Aufbau Suppressed Coupled Cluster Theory for Electronically Excited States
}

\author{Harrison Tuckman}
\affiliation{
Department of Chemistry, University of California, Berkeley, California 94720, USA 
}

\author{Eric Neuscamman}
\email{eneuscamman@berkeley.edu}
\affiliation{
Department of Chemistry, University of California, Berkeley, California 94720, USA 
}
\affiliation{Chemical Sciences Division, Lawrence Berkeley National Laboratory, Berkeley, CA, 94720, USA}

\date{\today}

\begin{abstract}
  We introduce an approach to improve single-reference coupled cluster theory
in settings where the Aufbau determinant is absent from or plays only
a small role in the true wave function.
Using a de-excitation operator that can be efficiently hidden within
a similarity transform, we create a coupled cluster wave function
in which de-excitations work to suppress the Aufbau determinant
and produce wave functions dominated by other determinants.
Thanks to an invertible and fully exponential form, the approach
is systematically improvable, size consistent, size extensive,
and, interestingly, size intensive in a granular way that should
make the adoption of some ground state techniques such
as local correlation relatively straightforward.
In this initial study, we apply the general formalism to create
a state-specific method for orbital-relaxed singly
excited states.
We find that this approach matches the accuracy of
similar-cost equation-of-motion methods in valence excitations
while offering improved accuracy for charge transfer states.
We also find the approach to be more accurate than
excited-state-specific perturbation theory in both types
of states.

\end{abstract}

\maketitle


\section{Introduction}

Coupled cluster (CC) theory
\cite{bartlett2007coupled,vcivzek1966correlation,vcivzek1971correlation,shavitt2009many}
offers highly accurate treatments of electron correlation and is particularly
effective in single-reference (SR) settings in which the true wave function is
dominated by a single determinant.
With strong formal properties like size extensivity and size consistency that
configuration interaction (CI) theory lacks,
\cite{helgaker2013molecular,crawford2007introduction}
CC can reliably achieve exquisite accuracy in SR settings,
as evidenced for example by
the sub-$k_B T$ errors of the perturbative-triples-corrected singles and doubles
theory, CCSD(T).
\cite{raghavachari1989fifth,watts1993coupled,thomas1993balance,helgaker1997prediction,bak2001accurate}
By looking a little closer at the details of CC theory, one can identify
two important factors driving its success in SR settings:
the accuracy of its reference and the conditioning of its working equations.
Indeed, when the true wave function is not dominated by a single determinant, SR-CC approaches often fail dramatically,
\cite{kohn2013state,lyakh2012multireference,lischka2018multireference}
and the construction of more multi-reference (MR) approaches
can lead to ill-conditioned equations
\textcolor{black}{or intruder state issues.}
\cite{van2001two,neuscamman2009quadratic,neuscamman2010strongly,neuscamman2010review,yanai2012extended,kowalski2000complete}

In thinking about MR-CC, it can be useful to recognize
that most approaches fall into one of three categories:
Jeziorski-Monkhorst (JM),
internal contraction,
and \textcolor{black}{single-reference-based methods}.
\cite{kohn2013state,lischka2018multireference,lyakh2012multireference}
\textcolor{black}{JM methods utilize a wave operator that contains a
separate cluster operator for each reference function within
the multi-reference starting point. \cite{jeziorski1981coupled}
In the state-universal (SU) approach, the idea is to
optimize this operator via a generalized Bloch equation
such that it transforms
linear combinations
of the reference functions
into a corresponding set of Hamiltonian eigenstates.
\cite{jeziorski1981coupled,piecuch1992orthogonally,paldus1993application,piecuch1994application,piecuch1994orthogonally,kucharski1991hilbert,balkova1991hilbert,balkova1991multi,balkova1994multireference,piecuch2002state}
In practice, SU approaches have often encountered intruder states,
unphysical solutions, and challenges in converging their equations,
\cite{paldus1993application,piecuch1994application,piecuch2002state,kowalski2000complete,kohn2013state}
in part because the
highest energy reference functions may not be energetically
well separated from determinants outside the reference
space. \cite{kohn2013state}
State-specific approaches to the JM ansatz avoid many of these
challenges, but, in eschewing the generalized Bloch equation,
lead to a situation in which there are more variables
than available projective equations, which in practice
has resulted in
the introduction of additional sufficiency conditions.}
\cite{hubavc1994size,pittner1999assessment,mahapatra2010potential,mahapatra2011evaluation,mahapatra1998state,mahapatra1999size,hanrath2005exponential}
Internally contracted approaches also employ multiple determinants
in the reference wave function,
but not in the construction of the cluster operator.
Instead, they act a single cluster operator on the reference,
which achieves a natural match between
the number of variables and projective equations
but creates challenges related to nonterminating expansions and ill-conditioned overlap matrices.
\cite{hanauer2011pilot,hanauer2012communication,evangelista2011orbital,datta2011state,li2004block}
\textcolor{black}{Single-reference-based methods,
on the other hand,} retain a single determinant reference but extend other parts
of the formalism to make it more amenable to MR settings.
\textcolor{black}{For example,
active-space-based approaches
include particular subsets of higher excitation operators, such as small subsets of the triples and quadruples.
\cite{piecuch1999coupled,oliphant1991multireference,piecuch1993state,piecuch2010active,shen2012combining,bauman2017combining,adamowicz2000new,lyakh2008generalization}
}

\textcolor{black}{These single-reference-based}
approaches retain many of the advantages of SR-CC
but also have challenges.
Like SR-CC, they have relatively straightforward working equations,
at least compared to those of MR-CC.
However, they can face difficulty in selecting the reference determinant,
for example in cases where the dominant determinant changes
along a reaction pathway. \cite{kohn2013state}
Further, a too-simple reference can necessitate large excitation amplitudes
that lead to large and unwelcome contributions
from the higher nonlinear terms in the CC expansion.
In the present study, we address the large amplitude issue directly
so that relatively simple SR working equations can be
usefully employed in a wider range of settings.

Specifically, we seek to avoid the need for
large excitation amplitudes even when the true
wave function has a small or zero Aufbau contribution
by including a de-excitation operator that
suppresses the Aufbau determinant within the CC expansion.
Possible applications of this Aufbau suppressed CC (ASCC) approach
include strongly correlated ground states as well as
state-specific treatments of both
weakly and strongly correlated excited states.
Unlike our previous approach of removing the Aufbau determinant
through a pseudo-projection operator, \cite{tuckman2023excited}
ASCC employs an invertible operator, making it
more general and avoiding the need to perturbatively correct
for projected-out pieces of the correlation treatment.
Although the ASCC formalism has the potential to be useful in
many areas, we focus in this initial study
on singly excited states, where it offers a route to
orbital-relaxed, excited-state-specific CC treatments atop
spin-pure reference functions.

One long-running challenge even in singly excited states has been
to fully capture the effects of post-excitation orbital relaxations.
\cite{subotnik2011communication}
Many widely used approaches, such as
time-dependent density functional theory (TD-DFT),
\cite{runge1984density,burke2005time,casida2012progress}
equation-of-motion CC (EOM-CC),
\cite{rowe1968equations,stanton1993equation,krylov2008equation}
and linear response CC (LR-CC),
\cite{monkhorst1977calculation,dalgaard1983some,sekino1984linear,koch1990excitation,koch1990coupled,rico1993single,koch1994calculation,sneskov2012excited}
rely on linear response and have only a limited ability to
account for orbital relaxations or the new correlation
effects created when an electron is transferred
between different regions of a molecule.
These issues are known to limit accuracy in
charge transfer (CT), Rydberg, core, and double excitations.
\cite{tozer1998improving,casida1998molecular,casida2000asymptotic,tozer2003importance,sobolewski2003ab,dreuw2003long,dreuw2004failure,mester2022charge,kozma2020new,izsak2020single}
State-specific approaches seek to improve matters
by more fully tailoring their treatments to the needs of the
excited state in question and
have been developed for single determinant theories,
\cite{ziegler1977calculation,kowalczyk2011assessment,gilbert2008self,besley2009self,barca2018simple,carter2020state,burton2020energy}
CI theory,
\cite{dreuw2005single,liu2012communication,liu2014variationally,shea2018communication,shea2020generalized,hardikar2020self,kossoski2022state,kossoski2023seniority,burton2022energy}
perturbation theory,
\cite{clune2020n5,clune2023studying}
CC theory,
\cite{mayhall2010multiple,lee2019excited,kossoski2021excited,marie2021variational,rishi2023dark,tuckman2023excited}
\textcolor{black}{complete active space self-consistent field (CASSCF)} theory,
\cite{knowles1985efficient,werner1985second,ruedenberg1982atoms,roos1987complete,tran2019tracking,tran2020improving,hanscam2022applying}
and DFT.
\cite{kowalczyk2011assessment,kowalczyk2013excitation,hait2020excited,hait2021orbital,zhao2019density,levi2020variational}
In this context, ASCC offers a route to systematically improvable,
state-specific CC
treatments with good spin symmetry and working equations that
closely mirror the ground state theory.

Through its fully exponential form, ASCC ensures its predictions are
size consistent, size extensive, and, perhaps most interestingly,
size intensive at a granular level.
\textcolor{black}{By granular, we mean that intensivity
is achieved not only in the
final result, but also at every stage within the optimization.
As we will see, this granular intensivity arises
because the mathematics for electrons far from the excitation simplify
to those of ground state SR-CC throughout the working equations.}
When combined with the use of excited-state-specific reference orbitals
from excited state mean field (ESMF) theory,
\cite{shea2018communication,shea2020generalized,hardikar2020self,clune2023studying}
this granular intensivity should make it relatively straightforward
to eventually incorporate local correlation treatments,
\cite{riplinger2013efficient,riplinger2013natural,riplinger2016sparse,saitow2017new,guo2018communication}
which may benefit from increased excited state specificity.
\cite{frank2018pair,helmich2011local,helmich2013pair}
In this initial study, however, we will focus on introducing the
general ASCC framework and investigating its
efficacy in singly excited states.

\section{Theory}

\subsection{A Challenge for Single-Reference Coupled Cluster}
\label{sec:sr_challenge}

In the standard single-reference approach, 
\cite{bartlett2007coupled,helgaker2013molecular,crawford2007introduction,shavitt2009many}
the coupled cluster (CC) equations are motivated by and derived from
an exponential ansatz
\begin{align}
  \ket{\Psi_{\mathrm{CC}}} = e^{\hat{T}}\ket{\phi _0}
            =\left(1+\hat{T}+\frac{1}{2}\hat{T}^2+\dots\right)\ket{\phi _0}\label{eqn: exponential}  
\end{align}
involving an excitation operator $\hat{T}$ and the closed-shell Aufbau determinant
$\ket{\phi_0}$.
If one allows $\hat{T}$ to contain all orders of excitation, then the theory is
exact within the one-electron basis, but in practice $\hat{T}$ is usually truncated
to produce a polynomial cost approach.
For example, CCSD includes only single and double excitations in $\hat{T}$.
Within the CI expansion that the CCSD wave function corresponds to,
this choice gives it the ability to set the coefficients on the singly and doubly
excited determinants however it wants while, unlike CI,
still achieving a size extensive energy.
Though CCSD has much more limited control over the coefficients on the triply
and higher excited determinants,
this is of little concern so long as two conditions are met:
(1) those highly excited terms are small in the exact wave function
and (2) the magnitude of $\hat{T}$ itself is small so that
its higher order powers do not create unduly large triply or higher
excited determinants.
When the first condition is not met (as is typically true in strongly
correlated systems), single-reference CC struggles. \cite{kohn2013state,cooper2010benchmark} 
However, what happens if one naively applies single-reference CC in
a case where only the second condition is not met?

\begin{table}[]
    \centering
    \begin{tabular}{c|c|rl}
       Determinant  & Label & \multicolumn{2}{c}{FCI Coefficient} \\
       \hline
        Aufbau                      & A&        $c_A=$ & 0.01 \\
        HOMO-X$\rightarrow$LUMO+X   & X & $\quad c_X=$ & 0.68 \\
        HOMO-Y$\rightarrow$LUMO+Y   & Y &       $c_Y=$ & 0.09 \\
        HOMO$^2\rightarrow$LUMO$^2$ & D &       $c_D=$ & 0.03 \\
        All others & & \multicolumn{2}{c}{very small}
    \end{tabular}
    \caption{Example of a simple singly excited state.}
    \label{tab:example_wfn}
\end{table}

For example, consider the relatively simple \textcolor{black}{full CI (FCI)} wave function
shown in Table \ref{tab:example_wfn}, which might arise
as one of a molecule's low-lying excited states.
Being dominated by single excitations, this state would
be well treated by many excited state theories, but
what would happen if we were to try to naively apply single-reference
CC to it directly, that is to say \textit{without} using
linear response theory?
Due to the fact that the CC expansion puts a coefficient of 1 on the
Aufbau determinant, the amplitude within $\hat{T}$ for the HOMO-X$\rightarrow$LUMO+X
single excitation would have to be huge --- something like $t_X\approx0.68/0.01=68$ ---
in order to get the right ratio of single-to-Aufbau in our expansion.
The others would have to be large as well:
$t_Y\approx0.09/0.01=9$ and $t_D\approx0.03/0.01=3$.
At this point, if we ignore normalization and look only through
linear order in the expansion, we'd be doing a good job
at matching FCI.
However, the nonlinear terms 
are now a disaster, as $\hat{T}$ is not small.
Consider the $\hat{T}^5$ term in the expansion, which leads to
a hextuply excited determinant with a coefficient on the order of
$t_X^2 t_Y^2 t_D / 5! \approx 9000$.
These types of terms now dominate the wave function, and so
instead of creating a good approximation of FCI, we've
created a terrible mess.
In sum, when trying to apply single-reference CC directly to a state in
which the Aufbau coefficient is small relative to some others,
we get in trouble regardless of whether we keep $\hat{T}$ small
or make it large.
If we keep it small, the Aufbau will be too large relative to the other
determinants, but, if we make it large, the nonlinear terms will
get out of hand.

This issue is why, of course, approaches other than a direct application
of single-reference CC are typically taken for this type of state.
Options include using a non-Aufbau reference, using multiple references,
and applying equation-of-motion or linear response theory.
Although these approaches all have their merits, and, depending on the details,
may work quite well for the example above, they also involve drawbacks.
The difficulty of capturing orbital relaxations in linear response theory
would be problematic if the state above were a charge transfer excitation.
Using an alternative single-determinant reference could work
well for the $c_X$ and $c_Y$ coefficients used in the example,
but would become challenging if molecular
geometry changes caused $c_X$ and $c_Y$ to gradually exchange magnitudes,
as it would then be difficult to choose which reference to use.
Were the $c_D$ coefficient to grow large, both of these approaches
would become more challenging.
At that point, one may be motivated to move away
from SR approaches, with all the complication
that that entails.
In this study, we will instead explore an alternative path that,
by suppressing the Aufbau determinant, provides a systematically
improvable and state-specific framework
in which a formal SR can be
maintained even when non-Aufbau determinants dominate the
true FCI wave function.

\subsection{Aufbau Suppression}

To suppress the coefficient on the Aufbau determinant in the expanded CC
wave function, let us augment our ansatz as follows.
\begin{align}
  \label{eqn:ascc_general_ansatz}\notag
  \ket{\Psi_{\mathrm{ASCC}}} &= e^{-\hat{S}^\dagger} e^{\hat{T}} \ket{\phi _0}\\
            &=\left( 1 + \hat{T} - \hat{S}^\dagger \hat{T}
                      + \frac{1}{2}\hat{T}^2 + \dots\right)\ket{\phi _0}
\end{align}
Here $\hat{S}$ is a (hopefully simple) excitation operator that will
be chosen such that the $-\hat{S}^\dagger \hat{T}\ket{\phi_0}$ term and any
similar higher-order terms produce additional copies of the Aufbau determinant
that can partially or fully cancel out the zeroth-order $\ket{\phi_0}$ term.
Basically, by de-exciting with $\hat{S}^\dagger$ after exciting with $\hat{T}$, we can
get back to Aufbau and, provided the amplitude coefficients within $\hat{S}$
and $\hat{T}$ are chosen carefully, can thereby suppress the overall coefficient
on the Aufbau determinant via cancellation.

Before discussing whether or not this approach can be practical, 
let us first emphasize its strong formal properties.
As the ansatz remains exponential, it will still product-factorize
and thus maintain size consistency.
Even better, so long as our approach to energy evaluation and optimization
does not introduce any unlinked terms, it will remain extensive.
The use of the invertible operator $\exp(-\hat{S}^\dagger)$ guarantees that
systematic improvability is also still present: if $\hat{T}$ contains
all orders of excitations, then $\exp(\hat{T})\ket{\phi_0}$
can describe $\exp(\hat{S}^\dagger)\ket{\Psi_{\mathrm{FCI}}}$
just as well as it can describe $\ket{\Psi_{\mathrm{FCI}}}$.
Finally, with the ability to suppress the Aufbau determinant, this approach
can describe wave functions dominated by non-Aufbau determinants
while keeping the magnitude of $\hat{T}$ modest, which should improve
our chances of achieving an accurate, state-specific
description while limiting $\hat{T}$ to low orders of excitation.

To make this approach practical, we first convert the central CC eigenvalue equation
\begin{align}
  \label{eqn:ascc_eig}
  \hat{H} \ket{\Psi_{\mathrm{ASCC}}} = E \ket{\Psi_{\mathrm{ASCC}}}
\end{align}
into a similarity transformed form in which $\hat{S}^\dagger$ has been
wrapped around the Hamiltonian.
\begin{align}
  \label{eqn:ascc_sim_eig}
  \bar{H} e^{\hat{T}} \ket{\phi_0} &= E e^{\hat{T}} \ket{\phi_0} \\
  \label{eqn:ascc_sim_ham}
  \bar{H} &= e^{\hat{S}^\dagger} \hat{H} e^{-\hat{S}^\dagger}
\end{align}
\textcolor{black}{Considering that extended CC (ECC) methods
induce large increases in computational cost 
when they introduce similar de-excitation-based transforms,
\cite{arponen1983variational,arponen1987extended,piecuch1999eomxcc,fan2005non,fan2006usefulness,van2000quadratic}
this step may seem counterproductive.
However, two key differences keep ASCC's cost in check.
First, the ordering of the excitation and de-excitation
operator exponentials is reversed compared to ECC, allowing
the de-excitation transform to act on the bare Hamiltonian.
Second, in ASCC, we restrict $\hat{S}$ to single excitations,
which allows Eq.\ (\ref{eqn:ascc_sim_ham}) to be evaluated
at $O(N^5)$ cost to yield a new set of one- and two-electron
integrals that, as we will see below, allow the overall method
to keep the same asymptotic scaling as ground state SR-CC.
Although in other settings it may be interesting to optimize
$\hat{S}$, we will now turn our attention to singly
excited states, where we will show that the desired
Aufbau suppression can be achieved with a particularly
simple and predetermined form.
}

\subsection{Singly Excited States}

Imagine a singly excited state in which one component of the excitation
is more important than the others, such as for example the wave function
in Table \ref{tab:example_wfn}.
To start with, let us consider the largest configuration state function
(CSF) within this wave function on its own before we get to
worrying about the smaller components or the weak correlation details.
We write this CSF as
\begin{equation}
    \ket{\psi _0}=\frac{1}{\sqrt{2}}\left(\ket{{\phi } _{h} ^{p}}+\ket{{\phi } _{\bar{h}} ^{\bar{p}}}\right) \label{eqn: truncated esmf}
\end{equation}
where the presence or absence of a bar on an index denotes opposite electron spin,
and $h$ and $p$ are the indices corresponding to the ``hole'' and ``particle''
spatial orbitals that are singly occupied in this CSF.
Adopting the terminology of the \textcolor{black}{active-space-based CC methods},
\cite{oliphant1991multireference}
we will refer to $\ket{\psi_0}$ as our \textit{reference}
(the major part of the state,
\textcolor{black}{Eq.\ (\ref{eqn: truncated esmf})},
around which we will fill in the details),
while we will refer to the Aufbau determinant
$\ket{\phi_0}$ as the \textit{formal reference}
(the state acted upon by $\exp(-\hat{S}^\dagger)\exp(\hat{T})$
in our ansatz definition \textcolor{black}{in Eq.\ (\ref{eqn:ascc_general_ansatz})}).
By choosing our operator $\hat{S}$ as the single excitation that excites
our formal reference to our reference (and whose adjoint de-excites the other way),
\begin{align}
    \label{eqn:1_CSF_S}
    \hat{S} \equiv \frac{1}{\sqrt{2}}
    &\left( \hat{a}^\dagger_p \hat{a}_h +\hat{a}^\dagger_{\bar{p}} \hat{a}_{\bar{h}} \right) \\
    \label{eqn:S_act_on_phi0}
    \hat{S}\ket{\phi_0}
    &= \ket{\psi_0} \\
    \label{eqn:S_adj_act_on_psi0}
    \hat{S}^\dagger \ket{\psi_0} &=
    \hat{S}^\dagger \hat{S} \ket{\phi_0} = \ket{\phi_0}
\end{align}
we set ourselves up to achieve Aufbau suppression via excitation to and
de-excitation from our reference CSF.
In particular, if we initialize our $\hat{T}$ operator to
the very simple singles and doubles form
\begin{align}
    \label{eqn:T_initial}
    \hat{T}_{\mathrm{init}} = \hat{S} - \frac{1}{2} \hat{S}^2
\end{align}
and note that $\hat{S}^3=0$,
\textcolor{black}{as one cannot
excite more than two electrons out of
the hole orbital,}
then our ASCC ansatz is initialized to our reference CSF $\ket{\psi_0}$.
\begin{align}
  e^{-\hat{S}^\dagger} e^{\hat{T}_{\mathrm{init}}} \ket{\phi_0}
  &= e^{-\hat{S}^\dagger}e^{\hat{S}-\frac{1}{2}\hat{S}^2}\ket{\phi_0} \\
  &= \left( 1 - \hat{S}^\dagger + \frac{1}{2} (\hat{S}^\dagger)^2 \right)
     \left(1+\hat{S}\right)\ket{\phi_0}\\
  &=  \left( 1 + \hat{S} - \hat{S}^\dagger \hat{S}  \right)\ket{\phi_0}\\
  &=  \hat{S} \ket{\phi_0}\\
  &=  \ket{\psi_0}
\end{align}
As intended, the Aufbau determinant has been suppressed, leaving us with a wave
function in which our reference CSF is dominant.
In other words, we have reached a qualitatively
correct starting point for describing many singly excited states.
This was achieved while maintaining the Aufbau determinant as our formal 
reference and while maintaining relatively modest amplitudes sizes within $\hat{T}$.
Indeed, the $1/\sqrt{2}$ and $-1/2$ amplitudes hiding inside
$\hat{T}_{\mathrm{init}}$ are much smaller than the values of 68 and 9
seen in our example in Section \ref{sec:sr_challenge}.
Whether they are small enough to allow for an accurate fleshing out
of the remaining wave function details is a matter for
numerical tests, which we explore in Section \ref{sec:results}.
While the attention thus far has been focused on states dominated by a single CSF, this formalism naturally extends to include states with multiple dominant CSFs as well by including additional excitation operators in $\hat{S}$. While increasing the number of excitation operators in $\hat{S}$ requires additional components in $\hat{T}_{\mathrm{init}}$ to exactly reconstruct the reference -- for example, a handful of triples amplitudes and a single quadruple amplitude in the two CSF case -- this subset of amplitudes remains relatively modest in size so long as the number of dominant CSFs in a state remains relatively small, as is often the case.
Having successfully built up our reference starting from our formal reference,
we now turn our attention to adding and optimizing the details needed for a
robust correlation treatment.

\subsection{Filling in the Details}

To think through which additional amplitudes should be enabled
to take us from $\hat{T}_{\mathrm{init}}$ to a $\hat{T}$
more in line with that used in CCSD,
it is useful to separate this operator into two pieces.
\begin{align}
\label{eqn:TP_TNP}
    \hat{T}=\hat{T}_{P}+\hat{T}_{NP}
\end{align}
\textcolor{black}{This partitioning is similar to the internal
and external
partitioning in single-reference-based multi-reference CC,
\cite{piecuch1993state,piecuch2010active}
although, for convenience of notation in this study,
we draw the line between the two sets of
amplitudes in a slightly different way.
Specifically, we} group into $\hat{T}_{P}$ all amplitudes whose excitation
operators contain one or more of our ``primary'' indices
\textcolor{black}{($h$, $p$, $\bar{h}$, $\bar{p}$) from}
within our chosen $\hat{S}$
from Eq.\ \ref{eqn:1_CSF_S}.
All other excitation operators will be grouped into the
``non-primary'' $\hat{T}_{NP}$.
\textcolor{black}{This way, since the terms in $\hat{T}_{NP}$
bear no primary indices while those in
$\hat{S}^\dagger$ bear only primary indices,
$\hat{T}_{NP}$ and $\hat{S}^\dagger$ will commute.
We can thus}
rearrange our ansatz as
\begin{align}
    \ket{\Psi}&=e^{\hat{T}_{NP}}e^{-\hat{S}^\dagger}e^{\hat{T}_P}\ket{\phi_0} \label{eqn: PandNP}
\end{align}
in which we see that the contributions from $\hat{T}_{NP}$ will behave much the
same as would those in multi-reference CC.
Indeed, $\exp(\hat{T}_{NP})$ can be formally understood as acting on the
Aufbau-suppressed, multi-determinant form set up by
$\exp(-\hat{S}^\dagger)\exp(\hat{T}_P)\ket{\phi_0}$, even if, operationally,
we group $\exp(-\hat{S}^\dagger)$ with $\hat{H}$ for computational expediency.
Recognizing this parallel to multi-reference CC and aiming for a
weak correlation treatment of similar quality to CCSD, we choose
$\hat{T}_{NP}$ to contain all single and double excitations that have
no primary indices.

In choosing what to include in $\hat{T}_{P}$, we want to ensure that we have
all singles and doubles relative to our reference $\ket{\psi_0}$.
Although this choice goes a bit beyond that, we start by including
all singles and doubles that have at least one primary index.
\textcolor{black}{However, some of the double excitations relative to our reference $\ket{\psi_0}$ that involve excitations from hole or to particle orbitals are actually triple excitations relative to our formal reference $\ket{\phi _0}$,\cite{tuckman2023excited} so, taking inspiration from the active-space-based CC methods,
\cite{piecuch1999coupled,oliphant1991multireference,piecuch1993state,piecuch2010active,shen2012combining,bauman2017combining,adamowicz2000new,lyakh2008generalization}}
we also include the
\textcolor{black}{small slice of triple excitations
$\hat{T}_{3^{'}}$}
that both
a) contain at least three primary indices and
b) contain a primary single excitation.
The inclusion of these triples is not a cost concern, as they are
only a small $O(ov^2)$ slice of all the triples.
Putting it all together, our excitation operator is
\begin{align}
    \label{eqn:amps_all_together}
    \hat{T}
    &= \hat{T}_{P} + \hat{T}_{NP}
    = \hat{T}_1 + \hat{T}_2 + \hat{T}_{3'}
\end{align}
\textcolor{black}{where $\hat{T}_1$ and $\hat{T}_2$
are the same as in CCSD and
$\hat{T}_{3'}$ is
the small slice of triples described above.}

One final detail to take care of is how, when appropriate, to allow some
contribution from the Aufbau determinant to survive.
One approach, which seems expedient especially when the contribution is
expected to be small, is to simply let the amplitude optimization modify
the $1/\sqrt{2}$ coefficients on the primary single excitation within
$\hat{T}_{\mathrm{init}}$ to values slightly above or below $1/\sqrt{2}$
so that the cancellation effect is no longer perfect and a small
contribution from Aufbau remains.
In the present study, this is the approach we will take.
We will mention, though, that if one wants the flexibility to
allow larger Aufbau contributions, one option would be to
simply scale down the $\hat{S}^\dagger$ operator via
\begin{equation}
\label{eqn: ASCC}
\ket{\Psi} =
e^{-(1-s_0)\hat{S}^\dagger}e^{\hat{T}}\ket{\phi _0}.
\end{equation}
Setting $s_0=1$ recovers standard SR-CC,
while values in between 0 and 1 provide varying amounts of Aufbau suppression.
Again, in this study we set $s_0=0$ and instead rely on the optimization of the
amplitudes in $\hat{T}$ to reintroduce Aufbau contributions where appropriate.

\subsection{Optimizing the Amplitudes} \label{sec: optimizing}

To determine the amplitudes within $\hat{T}$ and ultimately
evaluate the energy, we take the usual projective approach
of standard CC theory.
\cite{bartlett2007coupled,helgaker2013molecular,crawford2007introduction,shavitt2009many}  
\begin{align}
    \label{eqn: energy}
    E&
      = \bra{\phi_0 } e^{-\hat{T}} e^{\hat{S}^\dagger} \hat{H} e^{-\hat{S}^\dagger} e^{\hat{T}} \ket{\phi_0} \\
    \label{eqn: amplitudes}
    0&
      = \bra{\phi_\mu} e^{-\hat{T}} e^{\hat{S}^\dagger} \hat{H} e^{-\hat{S}^\dagger} e^{\hat{T}}\ket{\phi _0}
\end{align}
As in SR-CC, $\ket{\phi_\mu} = \hat{T}_{\mu} \ket{\phi_0}$
are the individual determinants reached by acting the individual
excitation operators $\hat{T}_{\mu}$ within $\hat{T}$ on the Aufbau determinant.
The value of using an invertible exponential form to achieve
Aufbau suppression becomes clear here, where it ensures that only connected
and therefore linked terms appear in the working equations, thereby guaranteeing size extensivity.
\cite{goldstone1957derivation,bartlett1978many,shavitt2009many} 
In practice, we set up our working equations by first performing
the MO integral transformation needed to form the
one- and two-electron integrals of the similarity transformed $\bar{H}$
from Eq.\ (\ref{eqn:ascc_sim_ham}), in terms of which our energy
and amplitude equations now directly mirror SR-CC theory.
\begin{align}
    \label{eqn:energy_barH}
    E&= \bra{\phi_0 } e^{-\hat{T}} \bar{H} e^{\hat{T}} \ket{\phi_0}  \\
    \label{eqn:amplitudes_barH}
    0&= \bra{\phi_\mu} e^{-\hat{T}} \bar{H} e^{\hat{T}}\ket{\phi_0} 
\end{align}
We should point out that some care in implementation is needed, as the two-electron integrals
within $\bar{H}$ lack the full 8-fold permutational symmetry that they had in $\hat{H}$, but importantly maintain the 4-fold symmetry necessary for grouping together terms as in the ground state theory.
Aside from that, the working equations are now $\textit{identical}$ to
those of a standard SR-CC theory in which $\hat{T}$ is chosen
according to Eq.\ (\ref{eqn:amps_all_together}).
Thanks to the inclusion of only a small slice of the triples,
these equations have the same $O(o^2v^4)$ asymptotic scaling as CCSD.

With working equation so similar to standard CC, we have for now simply followed
the standard recipe for solving the amplitude equations.
Specifically, we use a quasi-Newton iterative solver in which we approximate
the Jacobian as the diagonal of the similarity transformed Fock operator,
\textcolor{black}{which is obtained via a Fock-based analogue of Eq.\ \ref{eqn:ascc_sim_ham}. This update scheme closely parallels what is often done in ground
state SR-CC,\cite{helgaker2013molecular} and leads
to a familiar form for the amplitude updates.}
\begin{equation}
    t_\mu ^{new}=t_\mu^{old}-\frac{R_\mu}{\Delta _\mu}
\end{equation}
The residual $R_\mu$ is the right hand side of Eq.\ (\ref{eqn:amplitudes_barH}),
and $\Delta_\mu$ is the virtuals-minus-occupieds difference of the
similarity transformed Fock matrix diagonal entries corresponding to
the $\hat{T}_\mu$ excitation.
As we are not in the canonical basis, the Fock matrix is not diagonal,
\textcolor{black}{and so the diagonal Jacobian approximation is a somewhat more aggressive approximation than in canonical single-reference, ground state CC where the Fock matrix is diagonal}.
Nonetheless, when paired with
DIIS acceleration, \cite{pulay1980convergence}
this update scheme achieves tight convergence
($|R_\mu| < 10^{-10}$ a.u.) in all of our results.

As a final note regarding amplitude optimization, we should point out that ASCC
delivers what one might call iteration-by-iteration intensivity.
In addition to being size extensive, size consistent, and (for excited states)
size intensive thanks to its exponential ansatz,
the adoption of the same iterative solver used by the ground state
theory causes amplitudes on molecular fragments far away from the
fragment bearing the excitation to match those of CCSD at every step of the
optimization.
This essentially means that the ways in which the amplitude equations
differ from the ground state theory are ``localized'' around the excitation
itself.
Distant parts of the system experience essentially identical mathematics
as in CCSD, down to the level of individual amplitude updates.
We note that this property should make it relatively straightforward to adapt
many ground state techniques, such as local correlation methods,
for use in ASCC.

\subsection{Primary CSF and Orbital Basis}

So far, we have not specified how we choose the MO basis or the
primary CSF that we use to define $\hat{S}$ in this
single-excitation-focused incarnation of ASCC.
Although many choices are possible in principle,
including CASSCF,\textcolor{black}{\cite{knowles1985efficient,werner1985second,ruedenberg1982atoms,roos1987complete}} DFT,\textcolor{black}{\cite{hohenberg1964inhomogeneous,kohn1965self,parr1980density}} TD-DFT,\textcolor{black}{\cite{runge1984density,burke2005time,casida2012progress}} CIS,\textcolor{black}{\cite{dreuw2005single}} selected CI (sCI),\textcolor{black}{\cite{huron1973iterative,sharma2017semistochastic,garniron2018selected}}
and even EOM-CCSD,\textcolor{black}{\cite{rowe1968equations,stanton1993equation,krylov2008equation}}
in this study we  
elect to use excited state mean field theory (ESMF)\cite{shea2018communication,shea2020generalized,hardikar2020self} to generate an initial guess, because it provides a state specific, orbital relaxed, and spin-pure singly excited reference function. By transforming the ESMF wavefunction to its transition orbital pair (TOP) orbital basis,\cite{clune2020n5} which shares many similarities with the natural transition orbital (NTO) basis,\cite{martin2003natural} the full ESMF wavefunction is compressed into a smaller set of determinants without loss of information, which can then be truncated to include only the most important CSFs for an initial guess. 
Once we have truncated to these primary CSFs,
we separately re-canonicalize the non-primary parts
of the occupied and virtual spaces so as
to make the Fock matrix used in the
quasi-Newton amplitude update as close to
diagonal as possible without modifying
the truncated ESMF reference.
While this truncated ESMF serves its role relatively well for the singly excited states
in this initial study, the most effective methods
for generating ASCC's initial guess and orbital basis will likely
vary by application, and so it will be important to explore
the possibilities more systematically in the future.

\section{Results}
\label{sec:results}

\subsection{Computational Details}

For the single- and two-CSF QUEST tests,
the EOM-CCSD calculations were performed with PySCF,
\cite{sun2015libcint,sun2018pyscf,sun2020recent}
while for the charge transfer tests, EOM-CCSD and
$\delta$-CR-EOM-CC(2,3),A
\cite{piecuch2006single,piecuch2005renormalized,piecuch2015benchmarking,piecuch2002efficient,kowalski2004new,piecuch2009left,fradelos2011embedding}
calculations were performed with GAMESS.
\cite{schmidt1993general,barca2020recent}
These calculations, as well as ASCC calculations, did not use
the frozen core approximation.
We iterate ASCC until the maximum amplitude equation residual was no larger than $10^{-10}$, and for all other methods we utilize the default convergence settings. 
Geometries for the charge transfer tests were adapted from the cc-pVDZ geometries in the NIST Computational Chemistry Comparison and Benchmark Database \cite{johnson2022nist}
and can be found in the Supporting Information (SI).
Other geometries are from the QUEST \#1 excitation energy benchmark set\cite{loos2018mountaineering}. All calculations were performed in the aug-cc-pVDZ basis, with the exception that, in the charge transfer tests,
we removed augmentation from the hydrogen atoms. Any CSF with an ESMF singular value greater than 0.2 was included in the truncated ESMF reference for both ASCC and ESMP2.

\subsection{Single-CSF QUEST Tests}
\label{sec:single-CSF}

\setlength{\tabcolsep}{9pt}
\begin{table*}
    \footnotesize
    \centering
    \begin{tabular}{l l S[table-format=2.2] S[table-format=2.2] S[table-format=2.2] S[table-format=2.2]}
         Molecule & State            & {ESMP2} & {ASCC} & {EOM-CCSD} & {Reference$^{b}$} \\ \hline
            water & $1^1\text{B}_1$ &  7.59 &  7.50 &  7.45 &  7.53 \\                             
                  & $1^1\text{A}_2$ &  9.37 &  9.27 &  9.21 &  9.32 \\                             
                  & $2^1\text{A}_1$ &  9.95 &  9.86 &  9.86 &  9.94 \\                             
 hydrogen sulfide & $1^1\text{B}_1$ &  6.02 &  6.12 &  6.13 &  6.10 \\                             
                  & $1^1\text{A}_2$ &  6.08 &  6.28 &  6.34 &  6.29 \\                             
          ammonia & $1^1\text{A}_2$ &  6.41 &  6.42 &  6.46 &  6.48 \\                             
                  & $1^1\text{E}  $ &  8.05 &  8.03 &  8.03 &  8.08 \\                             
                  & $2^1\text{A}_1$ &  9.51 &  9.65 &  9.65 &  9.68 \\                             
                  & $2^1\text{A}_2$ & 10.29 & 10.45 & 10.38 & 10.41 \\                             
hydrogen chloride & $1^1\Pi$        &  7.71 &  7.82 &  7.86 &  7.82 \\                             
       dinitrogen & $1^1\Pi _g$     &  8.77 &  9.64 &  9.49 &  9.41 \\                             
  carbon monoxide & $1^1\Pi$        &  8.23 &  8.66 &  8.67 &  8.57 \\                             
                  & $2^1\Sigma ^+$  & 10.64 & 11.19 & 11.17 & 10.94 \\                             
                  & $3^1\Sigma ^+$  & 11.22 & 11.48 & 11.71 & 11.52 \\                             
                  & $2^1\Pi _u$     & 11.52 & 11.87 & 11.97 & 11.76 \\                             
         ethylene &  $1^1$B$_{3u}$  &  7.23 &  7.22 &  7.33 &  7.31 \\                             
                  &  $1^1$B$_{1u}$  &  7.79 &  7.88 &  8.04 &  7.93 \\                             
                  &  $1^1$B$_{1g}$  &  7.92 &  7.91 &  8.01 &  8.00 \\                             
     formaldehyde & $1^1\text{A}_2$ &  4.00 &  3.94 &  4.02 &  3.99 \\                             
                  & $1^1\text{B}_2$ &  7.35 &  7.11 &  7.04 &  7.11 \\                             
                  & $2^1\text{B}_2$ &  8.29 &  8.08 &  7.99 &  8.04 \\                             
                  & $2^1\text{A}_2$ &  8.90 &  8.72 &  8.61 &  8.65 \\                             
                  & $1^1\text{B}_1$ &  9.24 &  9.27 &  9.37 &  9.29 \\                             
 thioformaldehyde & $1^1\text{A}_2$ &  1.98 &  2.16 &  2.32 &  2.26$^{c}$ \\                             
                  & $1^1\text{B}_2$ &  5.85 &  5.85 &  5.84 &  5.83 \\                             
                  & $2^1\text{A}_1$ &  6.17 &  6.61 &  6.75 &  6.51 \\                             
      methanimine & $1^1\text{A}_d$ &  5.09 &  5.22 &  5.31 &  5.25 \\                             
     acetaldehyde & $1^1\text{A}_d$ &  4.33 &  4.30 &  4.36 &  4.34 \\                             
     cyclopropene & $1^1\text{B}_1$ &  6.36 &  6.77 &  6.78 &  6.71$^{d}$ \\                             
                  & $1^1\text{B}_2$ &  6.49 &  6.86 &  6.88 &  6.82 \\                             
     diazomethane & $1^1\text{A}_2$ &  2.72 &  2.97 &  3.23 &  3.09 \\                             
                  & $1^1\text{B}_1$ &  5.03 &  5.31 &  5.43 &  5.35 \\                             
                  & $2^1\text{A}_1$ &  5.29 &  5.84 &  5.90 &  5.79 \\                             
        formamide & $1^1\text{A}_d$ &  5.66 &  5.62 &  5.71 &  5.70\phantom{$^{c}$} \\                             
                  & $2^1\text{A}_p$ &  6.95 &  6.73 &  6.83 &  6.67 \\                             
                  & $4^1\text{A}_p$ &  7.57 &  7.40 &  7.41 &  7.29 \\                             
           ketene & $1^1\text{A}_2$ &  3.59 &  3.84 &  3.97 &  3.84 \\                             
                  & $1^1\text{B}_1$ &  5.68 &  5.93 &  5.94 &  5.88 \\                             
                  & $2^1\text{A}_2$ &  6.89 &  7.10 &  7.15 &  7.08 \\                             
   nitrosomethane & $1^1\text{A}_d$ &  2.02 &  2.04 &  2.00 &  1.99 \\                             
streptocyanine-C1 & $1^1\text{B}_2$ &  6.48 &  7.19 &  7.22 &  7.14 
    \end{tabular}
    \\
    \begin{tabular}{l l c c c c c}
\multicolumn{6}{c}{Single CSF Statistics}\\
& & {ESMP2} & {ASCC} & {EOM-CCSD} & \\\hline
\bf{MSE$^e$ $\pm$ Std. Dev.} & & -0.13$\pm$0.23  & 0.01 $\pm$ 0.08  &  0.05$\pm$0.09 & \\
\bf{MUE$^f$ $\pm$ Std. Dev.} & &  \phantom{-}0.21$\pm$0.16  &  0.06 $\pm$ 0.05  &  0.08$\pm$0.06 & \\
\bf{Max Error} &  & 0.66 & 0.25 & 0.24 & \\\\
    \end{tabular}
    \caption{Excitation Energies in eV for single-CSF states$^{a}$ from the QUEST benchmark.}
    \label{tab: quest}
    \raggedright
    $^{a}$States where one ESMF singular value is $>$0.2. $^{b}$The QUEST benchmark \cite{loos2018mountaineering} reference was exFCI unless noted otherwise. $^{c}$CCSDTQ reference. $^{d}$CCSDT reference. $^{e}$Mean signed error (MSE). $^{f}$Mean unsigned error (MUE).
\end{table*}

\begin{figure*}
    \centering
    \begin{subfigure}{0.8\linewidth}
        \includegraphics[width=\linewidth]{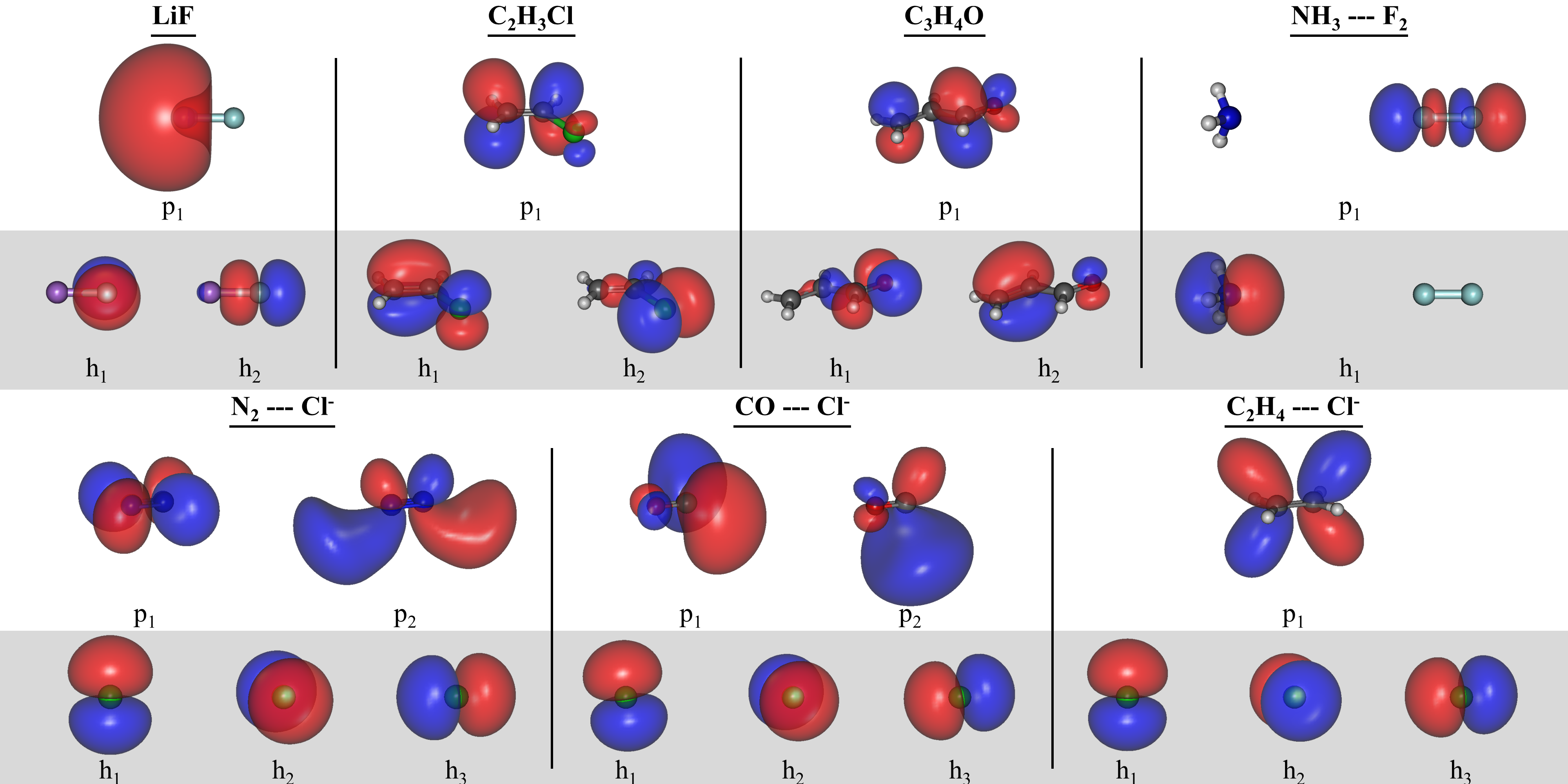}
        \caption{TOPs for particle and hole orbitals. Hole orbitals are in the grey background while particle orbitals are in the white. All possible combinations of hole and particle orbitals are considered. To save space, hole and particle orbitals are shown separately on chloride intermolecular systems.}
        \label{fig: orbitals}
    \end{subfigure}
    \begin{subfigure}{0.8\linewidth}
        \includegraphics[width=\linewidth,trim={0 3.8mm 0 0},clip]{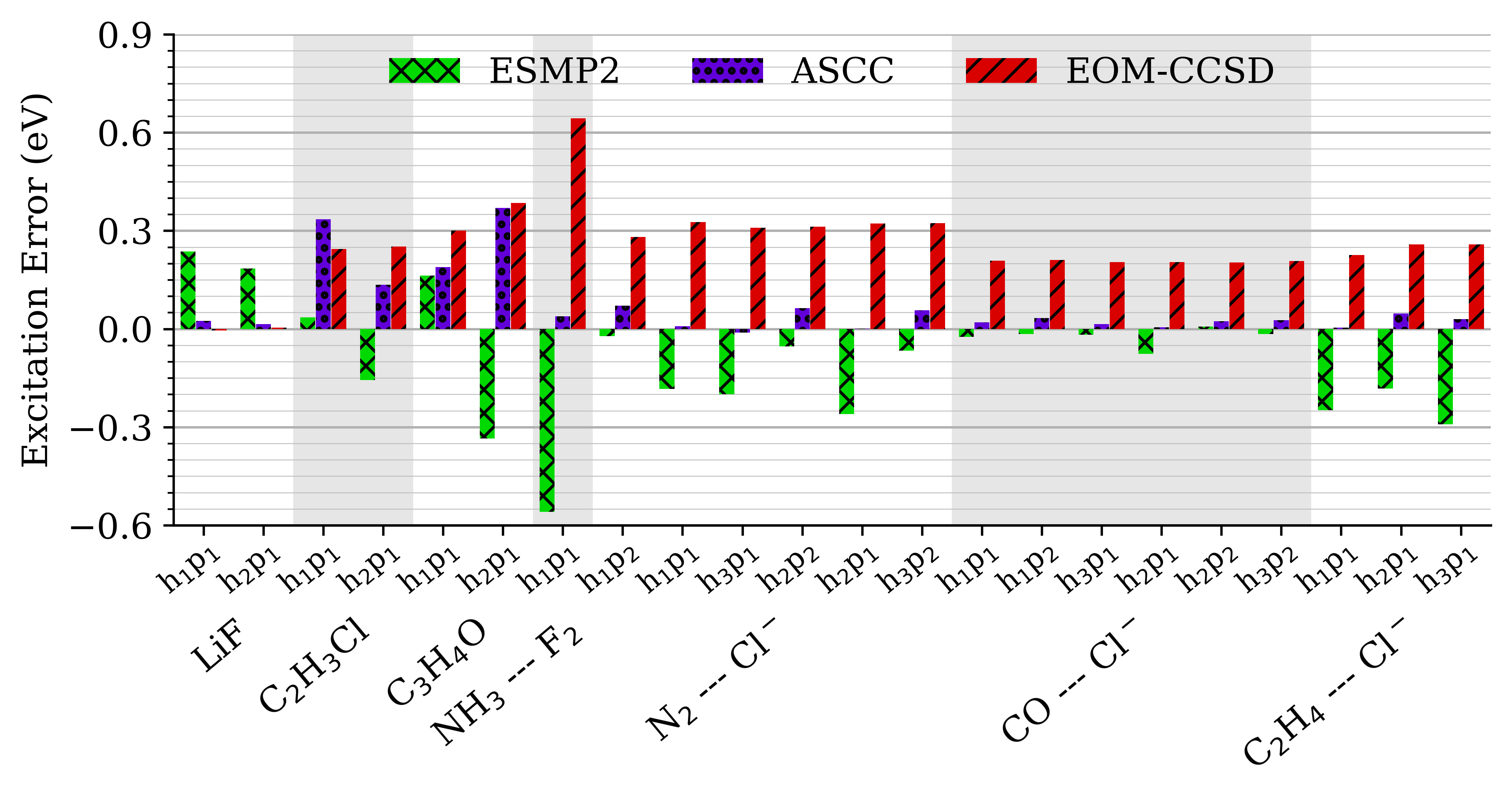}
        \caption{Excitation energy errors in eV relative to $\delta$-CR-EOM-CC(2,3),A.}
        \label{fig: ctplot}
    \end{subfigure}
    \begin{tabular}{l c c c}
        \multicolumn{4}{c}{Summary Statistics for Charge Transfer States}\\
        & {ESMP2} & {ASCC} & {EOM-CCSD} \\\hline
        \bf{MSE$^a$ $\pm$ Std. Dev.} & -0.09 $\pm$ 0.18 & 0.07 $\pm$ 0.10 & 0.26 $\pm$ 0.13 \\
        \bf{MUE$^b$ $\pm$ Std. Dev.} & \phantom{-}0.15 $\pm$ 0.14 & 0.07 $\pm$ 0.10 & 0.26 $\pm$ 0.13 \\
        \bf{Max Error} & 0.56 & 0.37 & 0.64 
    \end{tabular}\\
    $^a$Mean signed error (MSE). $^b$Mean unsigend error (MUE).
    \caption{Charge transfer orbitals (a), results (b), and summary statistics.}
    \label{fig: ct}
\end{figure*}

\begin{table*}
    \centering
    \footnotesize
    \begin{tabular}{l l S[table-format=2.2] S[table-format=2.2] S[table-format=2.2] S[table-format=2.2] S[table-format=2.2]}
         Molecule & State            & {ESMP2} & {ASCC} & {Frozen ASCC} & {EOM-CCSD} & {Reference$^{b}$} \\ \hline
        dinitrogen& $1^1\Sigma _u^-$&  9.78 & 10.90& 9.54 & 10.20 & 10.05 \\                             
                  & $1^1\Delta _u$  & 10.32 & 11.66& 10.34 & 10.61 & 10.43 \\                             
   carbon monoxide& $1^1\Sigma ^-$  &  9.92 & 11.12& 9.89 & 10.10 & 10.05 \\                             
                  & $1^1\Delta $    & 10.17 & 11.37& 10.08 & 10.21 & 10.16 \\                             
        acetylene &$1^1\Sigma _u ^-$&  6.83 &  7.99& 6.91 &  7.27 &  7.20 \\                             
                  & $1^1\Delta _u$  &  7.18 &  8.52& 7.43 &  7.57 &  7.51 \\                             
        formamide & $3^1\text{A}_p$ &  6.66 &  9.38& 8.06 &  7.72 &  7.64 \\                             
\phantom{streptocyanine-C1}  &  &  &  &  &  & 
    \end{tabular}
    \\
    \begin{tabular}{l l c c c c c}
\multicolumn{6}{c}{Multi CSF Statistics}\\
& & {ESMP2} & {ASCC} & {Frozen ASCC} & {EOM-CCSD} & \\\hline
\bf{MSE$^c$ $\pm$ Std. Dev.} & & -0.31 $\pm$ 0.32  & 1.13 $\pm$ 0.32 & -0.11 $\pm$ 0.28 & 0.09 $\pm$ 0.05 \\
\bf{MUE$^d$ $\pm$ Std. Dev.} & & \phantom{-}0.31 $\pm$ 0.32  & 1.13 $\pm$ 0.32  & \phantom{-}0.23 $\pm$ 0.18 & 0.09 $\pm$ 0.05\\
\bf{Max Error} & & 0.98 & 1.74 & 0.51 & 0.18 \\\\
    \end{tabular}
    \caption{Excitation Energies in eV for two-CSF states$^{a}$ from QUEST benchmark.}
    \label{tab: questcont}
    \raggedright
    \raggedright
    $^{a}$ States where two ESMF singular values are $>$0.2. $^{b}$ Reference from QUEST benchmark\cite{loos2018mountaineering} exFCI results unless stated otherwise. $^{c}$Mean signed error (MSE). $^{d}$Mean unsigned error (MUE).
\end{table*}

As an initial investigation of ASCC, we evaluate its accuracy on a
set of small molecule excitations from the QUEST \#1 excitation
energy benchmark. \cite{loos2018mountaineering}
Of the 56 total states, ESMF was able to provide a good initial guess
for 48 of them, with 8 others showing irreconcilable state mixing or
no convergence within ESMF.
Specifically, ESMF spuriously mixed large amounts of other
states into the Rydberg states of dinitrogen and the
totally symmetric excited states of formaldehyde,
while it did not converge for one singly and one doubly
excited state in nitrosomethane.
We have treated the 48 states with the present ASCC
approach, leaving the other 8
for future study with non-ESMF starting points.

Looking first at the states dominated by a single CSF,
the results for ASCC are compared to those of
EOM-CCSD, extrapolated FCI (exFCI),
high level CC, and ESMP2
\cite{shea2018communication,clune2020n5,clune2023studying}
(a state-specific 2$^{\mathrm{nd}}$ order perturbation theory
also built atop ESMF) in Table \ref{tab: quest}.
We note that the reference values make use of the
frozen core approximation while our calculations do not,
but the energetic differences resulting from this approximation
are rather small ($\sim$0.01 eV),\cite{loos2018mountaineering} so these excitation energies
still make excellent reference values. 

As one might expect, ASCC produces a mean unsigned error (MUE)
which is significantly lower than ESMP2's as well as a much smaller standard deviation.
Though the more robust correlation treatment in ASCC is likely responsible for the
majority of this difference, it is worth emphasizing ASCC's ability to adjust the
reference in the presence of the discovered details of the correlation.
Such adjustments can occur both through $\hat{T}_1$-based orbital relaxations
\cite{thouless1960stability}
as well as through adjustments to the size of other singly excited CSFs.
These abilities are both missing in the ESMP2 treatment and thus result in relatively large discrepancies between these two theories when the ESMF reference provides less reliable depictions of the target states, such as in the $3^1\Sigma ^+$ and $2^1 \Pi$ states of carbon monoxide where ESMF erroneously mixes in small amounts of
the $2^1 \Sigma ^+$ and $1^1 \Pi$ states respectively.
This issue appears to be responsible for ASCC's accuracy advantage
over ESMP2 in these states, where ASCC shows errors similar to those
in other states while ESMP2 shows unusually large errors.
This improved robustness against initial guess quality is a favorable
feature, especially considering that, in states where ESMF produces
an even less accurate mixing of CSFs, it could be necessary to
instead employ an alternative reference that may offer
less in the way of state-specific orbital relaxation.

Turning to a comparison with EOM-CCSD, ASCC produces a very slightly lower MUE,
but, given the statistical spreads, it's more correct to say that the two
methods showed about the same overall accuracy in these one-CSF QUEST tests.
The more salient difference appears to be that EOM-CCSD typically errors
slightly high,
while ASCC appears to error high and low in roughly equal amounts.
Given EOM-CCSD's well earned reputation for accuracy in small molecule
single excitations, it is encouraging to see comparable accuracy from ASCC
in such tests.
This observation in hand, we now turn to charge transfer, where
we would expect to see larger differences between state-specific and
linear response methods.

\subsection{Charge Transfer Tests}

To investigate ASCC in a context where post-excitation orbital relaxations
are expected to be more important, we applied it to a set of 
22 different charge transfer excitations, which consist of both intramolecular and intermolecular excitations characterized by significant dipole shifts.
Although we are currently limited to studying relatively small systems
by the pilot nature of our ASCC implementation, these states already
offer some clarity about the method's advantages in a CT context.
TOP orbital plots for each of these excitations are depicted in Figure \ref{fig: orbitals}, and excitation energy errors relative to $\delta$-CR-EOM-CC(2,3),A,
\cite{piecuch2006single,piecuch2005renormalized,piecuch2015benchmarking,piecuch2002efficient,kowalski2004new,piecuch2009left,fradelos2011embedding}
an $N^7$ scaling, size intensivity corrected, perturbative triples EOM-CC method, are shown in Figure \ref{fig: ctplot}.
Note that we elect to use the ``A'' variant of $\delta$-CR-EOM-CC(2,3) over others
due to its better-centered mean unsigned error. \cite{piecuch2015benchmarking}

\subsubsection{Intramolecular Charge Transfer}

Starting with lithium fluoride (LiF), we examine transfers into its
lithium 2s orbital (p$_1$)
from the fluorine off-axis (h$_1$) and on-axis (h$_2$) p orbitals.
In both of these excitations, ASCC and EOM-CCSD produce excitation energies in close alignment to the reference values, while ESMP2 produces somewhat larger errors. While EOM-CCSD often errs high for charge transfer excitations, in lithium fluoride, much like for hydrogen chloride in the QUEST set, the charge transfer distance is rather short and there are not too many electrons to
correlate.  Evidently, EOM-CCSD's linear response based approximations work well in this context.

In chloroethene (CH$_3$Cl), we examined transfers from both
the out-of-plane lone pair (h$_1$) and the in-plane lone pair
(h$_2$) to the $\pi^*$ orbital (p$_1$).
In these excitations, EOM-CCSD produces $\sim$0.25 eV errors,
which are larger than any of EOM-CCSD's errors in states considered
up till now.
For the h$_1$p$_1$ excitation, which has the smaller of the
two dipole shifts, ASCC errors a bit higher than EOM-CCSD,
while ESMP2 performs comparably to the reference method.
However, for the h$_2$p$_1$ excitation, whose dipole shift 
and CT character are more pronounced,
ASCC improves on EOM-CCSD with an error slightly smaller
than that of ESMP2.
We thus start to see what will become a pattern:
ASCC's performance relative to EOM-CCSD is better
in states with more pronounced CT character.

This pattern continues in acrolein (C$_3$H$_4$O),
where we examine transfers into a $\pi^*$ orbital (p$_1$)
from an oxygen lone pair (h$_1$) and from a
$\pi$ orbital (h$_2$).
The h$_1$p$_1$ excitation has the larger dipole shift and
more significant charge transfer character,
with the h$_1$ oxygen lone pair
delocalizing over the $\pi$ system.
ASCC's excitation energy error for h$_1$p$_1$ is less than
half the size of EOM-CCSD's, with ESMP2's error intermediate
between the two.
The h$_2$p$_1$ state has a smaller dipole shift and
less charge transfer character,
essentially shifting the electron within the $\pi$ system.
All three methods produce comparable and sizable errors
for this state, and although it is not entirely clear why
this is, it does continue the CT accuracy pattern
in which ASCC has an advantage when CT is more pronounced.

\subsubsection{Intermolecular Charge Transfer}

Turning now to states with even stronger CT character, we
find that ASCC has a clear accuracy advantage over EOM-CCSD
in intermolecular CT.
We begin with a transfer from ammonia to difluorine
in which the molecules are separated by 6 \r{A}.
The lowest-lying CT excitation in this system is from
the nitrogen lone pair (h$_1$) to difluorine's
$\sigma^*$ orbital (p$_1$).
EOM-CCSD's excitation energy error of 0.6 eV is its
largest error across all states tested in this study.
ESMP2 also produces a large error ($\sim$0.5 eV),
though it errors low while EOM-CCSD errors high.
ASCC, in contrast, produces an excitation energy in close agreement with the reference method.

As seen in Figure \ref{fig: ct}(b),
this pattern is repeated throughout our intermolecular CT tests:
EOM-CCSD errors high by 0.2 eV or more,
ESMP2 errors low by up to 0.3 eV,
and ASCC stays consistently within 0.1 eV (and usually 0.05 eV)
of the reference.
The lack of full orbital relaxation in EOM-CCSD likely explains most of
the difference, although we suspect that its neutral-ground-state-based
correlation treatment of the ionic CT state also plays a role.
Aside from NH$_3$ --- F$_2$, our intermolecular charge transfer
tests excite an electron from a chloride anion to a nearby
$\pi^*$ orbital of dinitrogen (N$_2$ --- Cl$^-$), carbon monoxide (CO --- Cl$^-$),
or ethylene (C$_2$H$_4$ --- Cl$^-$) at a separation distance of 4 \r{A}.
Some of these examples are artificial in that,
in a complete basis, they would only exist
as resonances,\cite{schulz1973resonances}
but in the finite basis we are using they still offer
meaningful tests of how closely methods come to
matching the energies of the $\delta$-CR-EOM-CC(2,3),A CT states.
The relatively large polarizability of the chlorine atom and
the transfer of a full electron across a significant distance
make orbital relaxation effects important in these states.
EOM-CCSD can only approximate these relaxations through its $\hat{R}_2$
operator, and its errors, which are significantly larger than
in the one-CSF QUEST tests of Section \ref{sec:single-CSF},
suggest that this approximation is reaching its limit.
In contrast, ASCC benefits from explicit orbital relaxations both from the ESMF reference
and its own $\hat{T}_1$ operator as well as from a correlation treatment optimized in the presence of the relaxed orbitals.

\subsection{Two-CSF QUEST Tests}

In this initial study of ASCC, we explore its performance in two-CSF
states via a minimal extension of the one-CSF approach in order to
determine whether further extensions are warranted.
In this minimal extension, \textcolor{black}{we begin by identifying the important hole and particle orbitals via the transformation of the ESMF wave function to its TOP orbital basis, just as in the 1-CSF case.}
We then update the $\hat{S}$ operator to
contain the four individual single excitation operators
(two alpha and two beta)
that create the two-CSF state, with their coefficients within $\hat{S}$
set by the ESMF wave function.
We extend $\hat{T}$ by adding the $O(1)$ subset of triples and the
single quadruple that flesh out a (4,4) active space encompassing
the two-CSF reference, and by extending the definition of a primary
orbital for the purposes of constructing $\hat{T}_{3'}$ to include
both CSFs' hole and particle orbitals.
We then set the initial amplitudes within $\hat{T}_{\mathrm{init}}$ so that
the initial ASCC wave function is equal to the two-CSF truncated ESMF state.

Optimizing this minimally extended two-CSF ASCC ansatz, we find that
further extensions will be needed in order for ASCC's systematic
improvability to produce accurate results in multi-CSF excited states.
Table \ref{tab: questcont} reveals errors on the order of 1 eV,
which stands in stark contrast to the more accurate results from
ESMP2 and EOM-CCSD.
Alongside these large energy errors, we see that some of
the initially nonzero amplitudes from $\hat{T}_{\mathrm{init}}$
--- which are important for creating the reference $\ket{\psi_0}$
from the Aufbau determinant $\ket{\phi_0}$ ---
change substantially during optimization.
To give a sense of scale, we can look at N$_2$,
where, during optimization, the sum of squares of the
primary CSFs' coefficients in the CI expansion of
the wave function
changed from 1.0 to 0.897 and 0.843 for the two-CSF
$1^1\Sigma^{-}_u$ and $1^1\Delta_u$ states, respectively,
which contrasts sharply with the change from 1.0 to 0.9995
seen in the one-CSF $1^1\Pi_g$ state,
which was typical of the other one-CSF states as well.
Given that the idea of a reference is that further changes should be
minor, this implies that either the reference was poor to begin with
or that artifacts arising from truncating $\hat{T}$ have led the
optimization to make erroneously large modifications to
these amplitudes.
With ESMP2 using the same reference and performing much better
than this minimal extension of
one-CSF ASCC, we think that the reference is not the issue.
Instead, we suspect that this minimal extension must be missing
components of the full $\hat{T}$ operator that were more important
than we had guessed.

What might these missing amplitudes be?  One possibility can be
spotted by noting that our minimal extension has created an ansatz
whose nonlinear terms will, if we expand the wave function out
into a CI basis, produce some strangely large quadruply excited
determinants.
For example, labeling our two primary excitations by the hole
and particle indices $h_1$, $h_2$, $p_1$, and $p_2$, the
$O(1)$-sized singles amplitude $t_{h_1}^{p_1}$ can combine
in a quadratic term with the $t_{\Bar{h}_1 h_2 k}^{\Bar{p}_1 b~c}$ 
triples amplitude to produce a quadruply excited determinant
with a coefficient roughly as large as that of a CCSD doubles
amplitude (because this triple is doubly excited relative to our reference).
There will be $O(ov^2)$ of these concerningly large quadruples,
and each of them contributes to the amplitude equation for
$t_{h_1 \Bar{h}_1}^{p_1 \Bar{p}_1}$, which plays an important
role in setting up the reference.
Were we to explicitly include the corresponding $O(ov^2)$
set of explicit quadruples within $\hat{T}$, this effect
would presumably be mitigated.
Although we do not think that this further extension would change
the theory's asymptotic scaling, we leave a more complete
study of how different extensions of ASCC's cluster operator
would improve accuracy in two- and multi-CSF states for future work.

Here, we attempt to answer the simpler question of whether
accuracy is improved by preventing the optimization of the weak
correlation part of the cluster operator from modifying the reference.
To do so, we break the optimization into two stages.
First, we only optimize the amplitudes within the (4,4) active space,
starting from the ESMF-based guess.
We then freeze those amplitudes and optimize the rest so as to prevent
the reference from being influenced by the spurious terms discussed above.
As seen in Table \ref{tab: questcont}, this ``Frozen ASCC'' approach
reduces errors considerably, although they are still larger than
in the single-CSF states.
These results suggest that, if the inclusion of small slices of
higher excitations within $\hat{T}$ prevents spurious
alterations to the reference, accuracy should be improved without
having to invoke a somewhat arbitrary freezing.

Thanks to ASCC's systematic improvability, we expect that adding key
higher amplitudes will ultimately lead to even greater
accuracy improvements.
Further, since the number of added amplitudes should be small,
we anticipate that they will not affect asymptotic scaling.
Finally, we note that adding these amplitudes will not create
mismatches between the number of amplitudes and equations,
as one can simply add the corresponding projections as well.
This contrasts with the need for sufficiency conditions
in \textcolor{black}{the state-specific }JM methods.
In the one-CSF regime, such conditions can be avoided via
two-determinant CC's spin-locking approach,
\cite{balkova1992coupled,lutz2018reference}
but it is not obvious how \textcolor{black}{these} methods would avoid them
in two-CSF states.

\section{Conclusion}

We have presented a new coupled cluster framework in which a de-excitation
operator is used to efficiently suppress the Aufbau
determinant within the wave function expansion.
This approach allows extended single-reference methods
to treat states in which the Aufbau determinant is small or absent
without resorting to large amplitude values and the problems they create.
Thanks to its fully exponential form, the approach is
systematically improvable, size consistent,
size extensive, and, for excited states, achieves a granular size intensivity
in which the equations for electrons far from the excitation simplify
into the ground state equations.
In initial testing on singly excited states, we find that a particularly
simple CCSD-like version of the theory works well in single-CSF
and especially charge transfer states, but that a more sophisticated
version will be needed for two-CSF states.

Looking forward, there are many exciting avenues to explore in
Aufbau suppressed coupled cluster.
An obvious direction is to determine which additional excitations
must be included to achieve accurate results for multi-CSF excited states.
Is it only the small slice of quadruples that we flagged above,
or will other amplitudes be important as well?
One could ask a similar question for doubly excited states,
which, if one chooses the MO basis the right way, can usually
be made to look like linear combinations of Aufbau, singles, and doubles.
In that case, as well as in some strongly correlated ground states,
it will be worth considering whether modifications to the de-excitation
operator would be helpful in addition to considering the inclusion
of higher excitation operators.
Other directions might seek to exploit the way that the theory's granular
size intensivity gives it a strong formal relationship with ground
state coupled cluster.
This connection could be helpful in adding local correlation treatments
and an analogue of the ground state's ``perturbative triples''
correction, since one can intuit that, apart from the amplitudes
directly involved in the excitation, these additions can be expected
to take a similar form as in the ground state.
In summary, we are excited to explore the possibilities created
by Aufbau suppression.

\section{Acknowledgements}

This work was supported by the National Science Foundation's
CAREER program under Award Number 1848012.
Calculations were performed 
using the Savio computational cluster resource provided by the Berkeley Research Computing program at the University of California, Berkeley and the Lawrencium computational cluster resource provided by the IT Division at the Lawrence Berkeley National Laboratory.
H.T. acknowledges that this 
material is based upon work supported by the National Science Foundation 
Graduate Research Fellowship Program under Grant No.\ 
DGE 2146752. Any opinions, findings, and conclusions or recommendations 
expressed in this material are those of the authors and do not necessarily 
reflect the views of the National Science Foundation.

\section{References}
\bibliographystyle{achemso}
\bibliography{main}

\providecommand{\latin}[1]{#1}
\makeatletter
\providecommand{\doi}
  {\begingroup\let\do\@makeother\dospecials
  \catcode`\{=1 \catcode`\}=2 \doi@aux}
\providecommand{\doi@aux}[1]{\endgroup\texttt{#1}}
\makeatother
\providecommand*\mcitethebibliography{\thebibliography}
\csname @ifundefined\endcsname{endmcitethebibliography}  {\let\endmcitethebibliography\endthebibliography}{}
\begin{mcitethebibliography}{154}
\providecommand*\natexlab[1]{#1}
\providecommand*\mciteSetBstSublistMode[1]{}
\providecommand*\mciteSetBstMaxWidthForm[2]{}
\providecommand*\mciteBstWouldAddEndPuncttrue
  {\def\EndOfBibitem{\unskip.}}
\providecommand*\mciteBstWouldAddEndPunctfalse
  {\let\EndOfBibitem\relax}
\providecommand*\mciteSetBstMidEndSepPunct[3]{}
\providecommand*\mciteSetBstSublistLabelBeginEnd[3]{}
\providecommand*\EndOfBibitem{}
\mciteSetBstSublistMode{f}
\mciteSetBstMaxWidthForm{subitem}{(\alph{mcitesubitemcount})}
\mciteSetBstSublistLabelBeginEnd
  {\mcitemaxwidthsubitemform\space}
  {\relax}
  {\relax}

\bibitem[Bartlett and Musia{\l}(2007)Bartlett, and Musia{\l}]{bartlett2007coupled}
Bartlett,~R.~J.; Musia{\l},~M. Coupled-cluster theory in quantum chemistry. \emph{Reviews of Modern Physics} \textbf{2007}, \emph{79}, 291\relax
\mciteBstWouldAddEndPuncttrue
\mciteSetBstMidEndSepPunct{\mcitedefaultmidpunct}
{\mcitedefaultendpunct}{\mcitedefaultseppunct}\relax
\EndOfBibitem
\bibitem[{\v{C}}{\'\i}{\v{z}}ek(1966)]{vcivzek1966correlation}
{\v{C}}{\'\i}{\v{z}}ek,~J. On the correlation problem in atomic and molecular systems. Calculation of wavefunction components in Ursell-type expansion using quantum-field theoretical methods. \emph{The Journal of Chemical Physics} \textbf{1966}, \emph{45}, 4256--4266\relax
\mciteBstWouldAddEndPuncttrue
\mciteSetBstMidEndSepPunct{\mcitedefaultmidpunct}
{\mcitedefaultendpunct}{\mcitedefaultseppunct}\relax
\EndOfBibitem
\bibitem[{\v{C}}{\'\i}{\v{z}}ek and Paldus(1971){\v{C}}{\'\i}{\v{z}}ek, and Paldus]{vcivzek1971correlation}
{\v{C}}{\'\i}{\v{z}}ek,~J.; Paldus,~J. Correlation problems in atomic and molecular systems III. Rederivation of the coupled-pair many-electron theory using the traditional quantum chemical methodst. \emph{International Journal of Quantum Chemistry} \textbf{1971}, \emph{5}, 359--379\relax
\mciteBstWouldAddEndPuncttrue
\mciteSetBstMidEndSepPunct{\mcitedefaultmidpunct}
{\mcitedefaultendpunct}{\mcitedefaultseppunct}\relax
\EndOfBibitem
\bibitem[Shavitt and Bartlett(2009)Shavitt, and Bartlett]{shavitt2009many}
Shavitt,~I.; Bartlett,~R.~J. \emph{Many-body methods in chemistry and physics: MBPT and coupled-cluster theory}; Cambridge university press, 2009\relax
\mciteBstWouldAddEndPuncttrue
\mciteSetBstMidEndSepPunct{\mcitedefaultmidpunct}
{\mcitedefaultendpunct}{\mcitedefaultseppunct}\relax
\EndOfBibitem
\bibitem[Helgaker \latin{et~al.}(2013)Helgaker, Jorgensen, and Olsen]{helgaker2013molecular}
Helgaker,~T.; Jorgensen,~P.; Olsen,~J. \emph{Molecular electronic-structure theory}; John Wiley \& Sons, 2013\relax
\mciteBstWouldAddEndPuncttrue
\mciteSetBstMidEndSepPunct{\mcitedefaultmidpunct}
{\mcitedefaultendpunct}{\mcitedefaultseppunct}\relax
\EndOfBibitem
\bibitem[Crawford and Schaefer~III(2007)Crawford, and Schaefer~III]{crawford2007introduction}
Crawford,~T.~D.; Schaefer~III,~H.~F. An introduction to coupled cluster theory for computational chemists. \emph{Reviews in computational chemistry} \textbf{2007}, \emph{14}, 33--136\relax
\mciteBstWouldAddEndPuncttrue
\mciteSetBstMidEndSepPunct{\mcitedefaultmidpunct}
{\mcitedefaultendpunct}{\mcitedefaultseppunct}\relax
\EndOfBibitem
\bibitem[Raghavachari \latin{et~al.}(1989)Raghavachari, Trucks, Pople, and Head-Gordon]{raghavachari1989fifth}
Raghavachari,~K.; Trucks,~G.~W.; Pople,~J.~A.; Head-Gordon,~M. A fifth-order perturbation comparison of electron correlation theories. \emph{Chemical Physics Letters} \textbf{1989}, \emph{157}, 479--483\relax
\mciteBstWouldAddEndPuncttrue
\mciteSetBstMidEndSepPunct{\mcitedefaultmidpunct}
{\mcitedefaultendpunct}{\mcitedefaultseppunct}\relax
\EndOfBibitem
\bibitem[Watts \latin{et~al.}(1993)Watts, Gauss, and Bartlett]{watts1993coupled}
Watts,~J.~D.; Gauss,~J.; Bartlett,~R.~J. Coupled-cluster methods with noniterative triple excitations for restricted open-shell Hartree--Fock and other general single determinant reference functions. Energies and analytical gradients. \emph{The Journal of Chemical Physics} \textbf{1993}, \emph{98}, 8718--8733\relax
\mciteBstWouldAddEndPuncttrue
\mciteSetBstMidEndSepPunct{\mcitedefaultmidpunct}
{\mcitedefaultendpunct}{\mcitedefaultseppunct}\relax
\EndOfBibitem
\bibitem[Thomas \latin{et~al.}(1993)Thomas, DeLeeuw, Vacek, Crawford, Yamaguchi, and Schaefer~III]{thomas1993balance}
Thomas,~J.~R.; DeLeeuw,~B.~J.; Vacek,~G.; Crawford,~T.~D.; Yamaguchi,~Y.; Schaefer~III,~H.~F. The balance between theoretical method and basis set quality: A systematic study of equilibrium geometries, dipole moments, harmonic vibrational frequencies, and infrared intensities. \emph{The Journal of chemical physics} \textbf{1993}, \emph{99}, 403--416\relax
\mciteBstWouldAddEndPuncttrue
\mciteSetBstMidEndSepPunct{\mcitedefaultmidpunct}
{\mcitedefaultendpunct}{\mcitedefaultseppunct}\relax
\EndOfBibitem
\bibitem[Helgaker \latin{et~al.}(1997)Helgaker, Gauss, J{\o}rgensen, and Olsen]{helgaker1997prediction}
Helgaker,~T.; Gauss,~J.; J{\o}rgensen,~P.; Olsen,~J. The prediction of molecular equilibrium structures by the standard electronic wave functions. \emph{The Journal of chemical physics} \textbf{1997}, \emph{106}, 6430--6440\relax
\mciteBstWouldAddEndPuncttrue
\mciteSetBstMidEndSepPunct{\mcitedefaultmidpunct}
{\mcitedefaultendpunct}{\mcitedefaultseppunct}\relax
\EndOfBibitem
\bibitem[Bak \latin{et~al.}(2001)Bak, Gauss, J{\o}rgensen, Olsen, Helgaker, and Stanton]{bak2001accurate}
Bak,~K.~L.; Gauss,~J.; J{\o}rgensen,~P.; Olsen,~J.; Helgaker,~T.; Stanton,~J.~F. The accurate determination of molecular equilibrium structures. \emph{The Journal of Chemical Physics} \textbf{2001}, \emph{114}, 6548--6556\relax
\mciteBstWouldAddEndPuncttrue
\mciteSetBstMidEndSepPunct{\mcitedefaultmidpunct}
{\mcitedefaultendpunct}{\mcitedefaultseppunct}\relax
\EndOfBibitem
\bibitem[K{\"o}hn \latin{et~al.}(2013)K{\"o}hn, Hanauer, Mueck, Jagau, and Gauss]{kohn2013state}
K{\"o}hn,~A.; Hanauer,~M.; Mueck,~L.~A.; Jagau,~T.-C.; Gauss,~J. State-specific multireference coupled-cluster theory. \emph{Wiley Interdisciplinary Reviews: Computational Molecular Science} \textbf{2013}, \emph{3}, 176--197\relax
\mciteBstWouldAddEndPuncttrue
\mciteSetBstMidEndSepPunct{\mcitedefaultmidpunct}
{\mcitedefaultendpunct}{\mcitedefaultseppunct}\relax
\EndOfBibitem
\bibitem[Lyakh \latin{et~al.}(2012)Lyakh, Musia{\l}, Lotrich, and Bartlett]{lyakh2012multireference}
Lyakh,~D.~I.; Musia{\l},~M.; Lotrich,~V.~F.; Bartlett,~R.~J. Multireference nature of chemistry: The coupled-cluster view. \emph{Chemical reviews} \textbf{2012}, \emph{112}, 182--243\relax
\mciteBstWouldAddEndPuncttrue
\mciteSetBstMidEndSepPunct{\mcitedefaultmidpunct}
{\mcitedefaultendpunct}{\mcitedefaultseppunct}\relax
\EndOfBibitem
\bibitem[Lischka \latin{et~al.}(2018)Lischka, Nachtigallova, Aquino, Szalay, Plasser, Machado, and Barbatti]{lischka2018multireference}
Lischka,~H.; Nachtigallova,~D.; Aquino,~A.~J.; Szalay,~P.~G.; Plasser,~F.; Machado,~F.~B.; Barbatti,~M. Multireference approaches for excited states of molecules. \emph{Chemical reviews} \textbf{2018}, \emph{118}, 7293--7361\relax
\mciteBstWouldAddEndPuncttrue
\mciteSetBstMidEndSepPunct{\mcitedefaultmidpunct}
{\mcitedefaultendpunct}{\mcitedefaultseppunct}\relax
\EndOfBibitem
\bibitem[Van~Voorhis and Head-Gordon(2001)Van~Voorhis, and Head-Gordon]{van2001two}
Van~Voorhis,~T.; Head-Gordon,~M. Two-body coupled cluster expansions. \emph{The Journal of Chemical Physics} \textbf{2001}, \emph{115}, 5033--5040\relax
\mciteBstWouldAddEndPuncttrue
\mciteSetBstMidEndSepPunct{\mcitedefaultmidpunct}
{\mcitedefaultendpunct}{\mcitedefaultseppunct}\relax
\EndOfBibitem
\bibitem[Neuscamman \latin{et~al.}(2009)Neuscamman, Yanai, and Chan]{neuscamman2009quadratic}
Neuscamman,~E.; Yanai,~T.; Chan,~G.~K. Quadratic canonical transformation theory and higher order density matrices. \emph{The Journal of chemical physics} \textbf{2009}, \emph{130}\relax
\mciteBstWouldAddEndPuncttrue
\mciteSetBstMidEndSepPunct{\mcitedefaultmidpunct}
{\mcitedefaultendpunct}{\mcitedefaultseppunct}\relax
\EndOfBibitem
\bibitem[Neuscamman \latin{et~al.}(2010)Neuscamman, Yanai, and Chan]{neuscamman2010strongly}
Neuscamman,~E.; Yanai,~T.; Chan,~G. K.-L. Strongly contracted canonical transformation theory. \emph{The Journal of chemical physics} \textbf{2010}, \emph{132}, 024106\relax
\mciteBstWouldAddEndPuncttrue
\mciteSetBstMidEndSepPunct{\mcitedefaultmidpunct}
{\mcitedefaultendpunct}{\mcitedefaultseppunct}\relax
\EndOfBibitem
\bibitem[Neuscamman \latin{et~al.}(2010)Neuscamman, Yanai, and Chan]{neuscamman2010review}
Neuscamman,~E.; Yanai,~T.; Chan,~G. K.-L. A review of canonical transformation theory. \emph{International Reviews in Physical Chemistry} \textbf{2010}, \emph{29}, 231--271\relax
\mciteBstWouldAddEndPuncttrue
\mciteSetBstMidEndSepPunct{\mcitedefaultmidpunct}
{\mcitedefaultendpunct}{\mcitedefaultseppunct}\relax
\EndOfBibitem
\bibitem[Yanai \latin{et~al.}(2012)Yanai, Kurashige, Neuscamman, and Chan]{yanai2012extended}
Yanai,~T.; Kurashige,~Y.; Neuscamman,~E.; Chan,~G. K.-L. Extended implementation of canonical transformation theory: parallelization and a new level-shifted condition. \emph{Physical Chemistry Chemical Physics} \textbf{2012}, \emph{14}, 7809--7820\relax
\mciteBstWouldAddEndPuncttrue
\mciteSetBstMidEndSepPunct{\mcitedefaultmidpunct}
{\mcitedefaultendpunct}{\mcitedefaultseppunct}\relax
\EndOfBibitem
\bibitem[Kowalski and Piecuch(2000)Kowalski, and Piecuch]{kowalski2000complete}
Kowalski,~K.; Piecuch,~P. Complete set of solutions of multireference coupled-cluster equations: The state-universal formalism. \emph{Physical Review A} \textbf{2000}, \emph{61}, 052506\relax
\mciteBstWouldAddEndPuncttrue
\mciteSetBstMidEndSepPunct{\mcitedefaultmidpunct}
{\mcitedefaultendpunct}{\mcitedefaultseppunct}\relax
\EndOfBibitem
\bibitem[Jeziorski and Monkhorst(1981)Jeziorski, and Monkhorst]{jeziorski1981coupled}
Jeziorski,~B.; Monkhorst,~H.~J. Coupled-cluster method for multideterminantal reference states. \emph{Physical Review A} \textbf{1981}, \emph{24}, 1668\relax
\mciteBstWouldAddEndPuncttrue
\mciteSetBstMidEndSepPunct{\mcitedefaultmidpunct}
{\mcitedefaultendpunct}{\mcitedefaultseppunct}\relax
\EndOfBibitem
\bibitem[Piecuch and Paldus(1992)Piecuch, and Paldus]{piecuch1992orthogonally}
Piecuch,~P.; Paldus,~J. Orthogonally spin-adapted multi-reference Hilbert space coupled-cluster formalism: Diagrammatic formulation. \emph{Theoretica chimica acta} \textbf{1992}, \emph{83}, 69--103\relax
\mciteBstWouldAddEndPuncttrue
\mciteSetBstMidEndSepPunct{\mcitedefaultmidpunct}
{\mcitedefaultendpunct}{\mcitedefaultseppunct}\relax
\EndOfBibitem
\bibitem[Paldus \latin{et~al.}(1993)Paldus, Piecuch, Pylypow, and Jeziorski]{paldus1993application}
Paldus,~J.; Piecuch,~P.; Pylypow,~L.; Jeziorski,~B. Application of Hilbert-space coupled-cluster theory to simple (H 2) 2 model systems: Planar models. \emph{Physical Review A} \textbf{1993}, \emph{47}, 2738\relax
\mciteBstWouldAddEndPuncttrue
\mciteSetBstMidEndSepPunct{\mcitedefaultmidpunct}
{\mcitedefaultendpunct}{\mcitedefaultseppunct}\relax
\EndOfBibitem
\bibitem[Piecuch and Paldus(1994)Piecuch, and Paldus]{piecuch1994application}
Piecuch,~P.; Paldus,~J. Application of Hilbert-space coupled-cluster theory to simple (H 2) 2 model systems. II. Nonplanar models. \emph{Physical Review A} \textbf{1994}, \emph{49}, 3479\relax
\mciteBstWouldAddEndPuncttrue
\mciteSetBstMidEndSepPunct{\mcitedefaultmidpunct}
{\mcitedefaultendpunct}{\mcitedefaultseppunct}\relax
\EndOfBibitem
\bibitem[Piecuch and Paldus(1994)Piecuch, and Paldus]{piecuch1994orthogonally}
Piecuch,~P.; Paldus,~J. Orthogonally spin-adapted state-universal coupled-cluster formalism: Implementation of the complete two-reference theory including cubic and quartic coupling terms. \emph{The Journal of chemical physics} \textbf{1994}, \emph{101}, 5875--5890\relax
\mciteBstWouldAddEndPuncttrue
\mciteSetBstMidEndSepPunct{\mcitedefaultmidpunct}
{\mcitedefaultendpunct}{\mcitedefaultseppunct}\relax
\EndOfBibitem
\bibitem[Kucharski and Bartlett(1991)Kucharski, and Bartlett]{kucharski1991hilbert}
Kucharski,~S.~A.; Bartlett,~R.~J. Hilbert space multireference coupled-cluster methods. I. The single and double excitation model. \emph{The Journal of chemical physics} \textbf{1991}, \emph{95}, 8227--8238\relax
\mciteBstWouldAddEndPuncttrue
\mciteSetBstMidEndSepPunct{\mcitedefaultmidpunct}
{\mcitedefaultendpunct}{\mcitedefaultseppunct}\relax
\EndOfBibitem
\bibitem[Balkov{\'a} \latin{et~al.}(1991)Balkov{\'a}, Kucharski, Meissner, and Bartlett]{balkova1991hilbert}
Balkov{\'a},~A.; Kucharski,~S.; Meissner,~L.; Bartlett,~R.~J. A Hilbert space multi-reference coupled-cluster study of the H 4 model system. \emph{Theoretica chimica acta} \textbf{1991}, \emph{80}, 335--348\relax
\mciteBstWouldAddEndPuncttrue
\mciteSetBstMidEndSepPunct{\mcitedefaultmidpunct}
{\mcitedefaultendpunct}{\mcitedefaultseppunct}\relax
\EndOfBibitem
\bibitem[Balkov{\'a} \latin{et~al.}(1991)Balkov{\'a}, Kucharski, and Bartlett]{balkova1991multi}
Balkov{\'a},~A.; Kucharski,~S.; Bartlett,~R.~J. The multi-reference Hilbert space coupled-cluster study of the Li2 molecule. Application in a complete model space. \emph{Chemical physics letters} \textbf{1991}, \emph{182}, 511--518\relax
\mciteBstWouldAddEndPuncttrue
\mciteSetBstMidEndSepPunct{\mcitedefaultmidpunct}
{\mcitedefaultendpunct}{\mcitedefaultseppunct}\relax
\EndOfBibitem
\bibitem[Balkov{\'a} and Bartlett(1994)Balkov{\'a}, and Bartlett]{balkova1994multireference}
Balkov{\'a},~A.; Bartlett,~R.~J. A multireference coupled-cluster study of the ground state and lowest excited states of cyclobutadiene. \emph{The Journal of chemical physics} \textbf{1994}, \emph{101}, 8972--8987\relax
\mciteBstWouldAddEndPuncttrue
\mciteSetBstMidEndSepPunct{\mcitedefaultmidpunct}
{\mcitedefaultendpunct}{\mcitedefaultseppunct}\relax
\EndOfBibitem
\bibitem[Piecuch and Kowalski(2002)Piecuch, and Kowalski]{piecuch2002state}
Piecuch,~P.; Kowalski,~K. The state-universal multi-reference coupled-cluster theory: An overview of some recent advances. \emph{International Journal of Molecular Sciences} \textbf{2002}, \emph{3}, 676--709\relax
\mciteBstWouldAddEndPuncttrue
\mciteSetBstMidEndSepPunct{\mcitedefaultmidpunct}
{\mcitedefaultendpunct}{\mcitedefaultseppunct}\relax
\EndOfBibitem
\bibitem[Huba{\v{c}} and Neogr{\'a}dy(1994)Huba{\v{c}}, and Neogr{\'a}dy]{hubavc1994size}
Huba{\v{c}},~I.; Neogr{\'a}dy,~P. Size-consistent Brillouin-Wigner perturbation theory with an exponentially parametrized wave function: Brillouin-Wigner coupled-cluster theory. \emph{Physical Review A} \textbf{1994}, \emph{50}, 4558\relax
\mciteBstWouldAddEndPuncttrue
\mciteSetBstMidEndSepPunct{\mcitedefaultmidpunct}
{\mcitedefaultendpunct}{\mcitedefaultseppunct}\relax
\EndOfBibitem
\bibitem[Pittner \latin{et~al.}(1999)Pittner, Nachtigall, {\v{C}}{\'a}rsky, M{\'a}{\v{s}}ik, and Huba{\v{c}}]{pittner1999assessment}
Pittner,~J.; Nachtigall,~P.; {\v{C}}{\'a}rsky,~P.; M{\'a}{\v{s}}ik,~J.; Huba{\v{c}},~I. Assessment of the single-root multireference Brillouin--Wigner coupled-cluster method: Test calculations on CH 2, SiH 2, and twisted ethylene. \emph{The Journal of chemical physics} \textbf{1999}, \emph{110}, 10275--10282\relax
\mciteBstWouldAddEndPuncttrue
\mciteSetBstMidEndSepPunct{\mcitedefaultmidpunct}
{\mcitedefaultendpunct}{\mcitedefaultseppunct}\relax
\EndOfBibitem
\bibitem[Mahapatra and Chattopadhyay(2010)Mahapatra, and Chattopadhyay]{mahapatra2010potential}
Mahapatra,~U.~S.; Chattopadhyay,~S. Potential energy surface studies via a single root multireference coupled cluster theory. \emph{The Journal of chemical physics} \textbf{2010}, \emph{133}, 074102\relax
\mciteBstWouldAddEndPuncttrue
\mciteSetBstMidEndSepPunct{\mcitedefaultmidpunct}
{\mcitedefaultendpunct}{\mcitedefaultseppunct}\relax
\EndOfBibitem
\bibitem[Mahapatra and Chattopadhyay(2011)Mahapatra, and Chattopadhyay]{mahapatra2011evaluation}
Mahapatra,~U.~S.; Chattopadhyay,~S. Evaluation of the performance of single root multireference coupled cluster method for ground and excited states, and its application to geometry optimization. \emph{The Journal of Chemical Physics} \textbf{2011}, \emph{134}, 044113\relax
\mciteBstWouldAddEndPuncttrue
\mciteSetBstMidEndSepPunct{\mcitedefaultmidpunct}
{\mcitedefaultendpunct}{\mcitedefaultseppunct}\relax
\EndOfBibitem
\bibitem[Mahapatra \latin{et~al.}(1998)Mahapatra, Datta, and Mukherjee]{mahapatra1998state}
Mahapatra,~U.~S.; Datta,~B.; Mukherjee,~D. A state-specific multi-reference coupled cluster formalism with molecular applications. \emph{Molecular Physics} \textbf{1998}, \emph{94}, 157--171\relax
\mciteBstWouldAddEndPuncttrue
\mciteSetBstMidEndSepPunct{\mcitedefaultmidpunct}
{\mcitedefaultendpunct}{\mcitedefaultseppunct}\relax
\EndOfBibitem
\bibitem[Mahapatra \latin{et~al.}(1999)Mahapatra, Datta, and Mukherjee]{mahapatra1999size}
Mahapatra,~U.~S.; Datta,~B.; Mukherjee,~D. A size-consistent state-specific multireference coupled cluster theory: Formal developments and molecular applications. \emph{The Journal of chemical physics} \textbf{1999}, \emph{110}, 6171--6188\relax
\mciteBstWouldAddEndPuncttrue
\mciteSetBstMidEndSepPunct{\mcitedefaultmidpunct}
{\mcitedefaultendpunct}{\mcitedefaultseppunct}\relax
\EndOfBibitem
\bibitem[Hanrath(2005)]{hanrath2005exponential}
Hanrath,~M. An exponential multireference wave-function Ansatz. \emph{The Journal of chemical physics} \textbf{2005}, \emph{123}, 084102\relax
\mciteBstWouldAddEndPuncttrue
\mciteSetBstMidEndSepPunct{\mcitedefaultmidpunct}
{\mcitedefaultendpunct}{\mcitedefaultseppunct}\relax
\EndOfBibitem
\bibitem[Hanauer and K{\"o}hn(2011)Hanauer, and K{\"o}hn]{hanauer2011pilot}
Hanauer,~M.; K{\"o}hn,~A. Pilot applications of internally contracted multireference coupled cluster theory, and how to choose the cluster operator properly. \emph{The Journal of chemical physics} \textbf{2011}, \emph{134}, 204111\relax
\mciteBstWouldAddEndPuncttrue
\mciteSetBstMidEndSepPunct{\mcitedefaultmidpunct}
{\mcitedefaultendpunct}{\mcitedefaultseppunct}\relax
\EndOfBibitem
\bibitem[Hanauer and K{\"o}hn(2012)Hanauer, and K{\"o}hn]{hanauer2012communication}
Hanauer,~M.; K{\"o}hn,~A. Communication: Restoring full size extensivity in internally contracted multireference coupled cluster theory. \emph{The Journal of Chemical Physics} \textbf{2012}, \emph{137}, 131103\relax
\mciteBstWouldAddEndPuncttrue
\mciteSetBstMidEndSepPunct{\mcitedefaultmidpunct}
{\mcitedefaultendpunct}{\mcitedefaultseppunct}\relax
\EndOfBibitem
\bibitem[Evangelista and Gauss(2011)Evangelista, and Gauss]{evangelista2011orbital}
Evangelista,~F.~A.; Gauss,~J. An orbital-invariant internally contracted multireference coupled cluster approach. \emph{The Journal of chemical physics} \textbf{2011}, \emph{134}, 114102\relax
\mciteBstWouldAddEndPuncttrue
\mciteSetBstMidEndSepPunct{\mcitedefaultmidpunct}
{\mcitedefaultendpunct}{\mcitedefaultseppunct}\relax
\EndOfBibitem
\bibitem[Datta \latin{et~al.}(2011)Datta, Kong, and Nooijen]{datta2011state}
Datta,~D.; Kong,~L.; Nooijen,~M. A state-specific partially internally contracted multireference coupled cluster approach. \emph{The Journal of chemical physics} \textbf{2011}, \emph{134}, 214116\relax
\mciteBstWouldAddEndPuncttrue
\mciteSetBstMidEndSepPunct{\mcitedefaultmidpunct}
{\mcitedefaultendpunct}{\mcitedefaultseppunct}\relax
\EndOfBibitem
\bibitem[Li(2004)]{li2004block}
Li,~S. Block-correlated coupled cluster theory: The general formulation and its application to the antiferromagnetic Heisenberg model. \emph{The Journal of chemical physics} \textbf{2004}, \emph{120}, 5017--5026\relax
\mciteBstWouldAddEndPuncttrue
\mciteSetBstMidEndSepPunct{\mcitedefaultmidpunct}
{\mcitedefaultendpunct}{\mcitedefaultseppunct}\relax
\EndOfBibitem
\bibitem[Piecuch \latin{et~al.}(1999)Piecuch, Kucharski, and Bartlett]{piecuch1999coupled}
Piecuch,~P.; Kucharski,~S.~A.; Bartlett,~R.~J. Coupled-cluster methods with internal and semi-internal triply and quadruply excited clusters: CCSD t and CCSD tq approaches. \emph{The Journal of chemical physics} \textbf{1999}, \emph{110}, 6103--6122\relax
\mciteBstWouldAddEndPuncttrue
\mciteSetBstMidEndSepPunct{\mcitedefaultmidpunct}
{\mcitedefaultendpunct}{\mcitedefaultseppunct}\relax
\EndOfBibitem
\bibitem[Oliphant and Adamowicz(1991)Oliphant, and Adamowicz]{oliphant1991multireference}
Oliphant,~N.; Adamowicz,~L. Multireference coupled-cluster method using a single-reference formalism. \emph{The Journal of chemical physics} \textbf{1991}, \emph{94}, 1229--1235\relax
\mciteBstWouldAddEndPuncttrue
\mciteSetBstMidEndSepPunct{\mcitedefaultmidpunct}
{\mcitedefaultendpunct}{\mcitedefaultseppunct}\relax
\EndOfBibitem
\bibitem[Piecuch \latin{et~al.}(1993)Piecuch, Oliphant, and Adamowicz]{piecuch1993state}
Piecuch,~P.; Oliphant,~N.; Adamowicz,~L. A state-selective multireference coupled-cluster theory employing the single-reference formalism. \emph{The Journal of chemical physics} \textbf{1993}, \emph{99}, 1875--1900\relax
\mciteBstWouldAddEndPuncttrue
\mciteSetBstMidEndSepPunct{\mcitedefaultmidpunct}
{\mcitedefaultendpunct}{\mcitedefaultseppunct}\relax
\EndOfBibitem
\bibitem[Piecuch(2010)]{piecuch2010active}
Piecuch,~P. Active-space coupled-cluster methods. \emph{Molecular Physics} \textbf{2010}, \emph{108}, 2987--3015\relax
\mciteBstWouldAddEndPuncttrue
\mciteSetBstMidEndSepPunct{\mcitedefaultmidpunct}
{\mcitedefaultendpunct}{\mcitedefaultseppunct}\relax
\EndOfBibitem
\bibitem[Shen and Piecuch(2012)Shen, and Piecuch]{shen2012combining}
Shen,~J.; Piecuch,~P. Combining active-space coupled-cluster methods with moment energy corrections via the CC (P; Q) methodology, with benchmark calculations for biradical transition states. \emph{The Journal of chemical physics} \textbf{2012}, \emph{136}, 144104\relax
\mciteBstWouldAddEndPuncttrue
\mciteSetBstMidEndSepPunct{\mcitedefaultmidpunct}
{\mcitedefaultendpunct}{\mcitedefaultseppunct}\relax
\EndOfBibitem
\bibitem[Bauman \latin{et~al.}(2017)Bauman, Shen, and Piecuch]{bauman2017combining}
Bauman,~N.~P.; Shen,~J.; Piecuch,~P. Combining active-space coupled-cluster approaches with moment energy corrections via the CC (P; Q) methodology: connected quadruple excitations. \emph{Molecular Physics} \textbf{2017}, \emph{115}, 2860--2891\relax
\mciteBstWouldAddEndPuncttrue
\mciteSetBstMidEndSepPunct{\mcitedefaultmidpunct}
{\mcitedefaultendpunct}{\mcitedefaultseppunct}\relax
\EndOfBibitem
\bibitem[Adamowicz \latin{et~al.}(2000)Adamowicz, Malrieu, and Ivanov]{adamowicz2000new}
Adamowicz,~L.; Malrieu,~J.-P.; Ivanov,~V.~V. New approach to the state-specific multireference coupled-cluster formalism. \emph{The Journal of Chemical Physics} \textbf{2000}, \emph{112}, 10075--10084\relax
\mciteBstWouldAddEndPuncttrue
\mciteSetBstMidEndSepPunct{\mcitedefaultmidpunct}
{\mcitedefaultendpunct}{\mcitedefaultseppunct}\relax
\EndOfBibitem
\bibitem[Lyakh \latin{et~al.}(2008)Lyakh, Ivanov, and Adamowicz]{lyakh2008generalization}
Lyakh,~D.~I.; Ivanov,~V.~V.; Adamowicz,~L. A generalization of the state-specific complete-active-space coupled-cluster method for calculating electronic excited states. \emph{The Journal of chemical physics} \textbf{2008}, \emph{128}\relax
\mciteBstWouldAddEndPuncttrue
\mciteSetBstMidEndSepPunct{\mcitedefaultmidpunct}
{\mcitedefaultendpunct}{\mcitedefaultseppunct}\relax
\EndOfBibitem
\bibitem[Tuckman and Neuscamman(2023)Tuckman, and Neuscamman]{tuckman2023excited}
Tuckman,~H.; Neuscamman,~E. Excited-State-Specific Pseudoprojected Coupled-Cluster Theory. \emph{Journal of Chemical Theory and Computation} \textbf{2023}, \relax
\mciteBstWouldAddEndPunctfalse
\mciteSetBstMidEndSepPunct{\mcitedefaultmidpunct}
{}{\mcitedefaultseppunct}\relax
\EndOfBibitem
\bibitem[Subotnik(2011)]{subotnik2011communication}
Subotnik,~J.~E. Communication: Configuration interaction singles has a large systematic bias against charge-transfer states. \emph{The Journal of chemical physics} \textbf{2011}, \emph{135}, 071104\relax
\mciteBstWouldAddEndPuncttrue
\mciteSetBstMidEndSepPunct{\mcitedefaultmidpunct}
{\mcitedefaultendpunct}{\mcitedefaultseppunct}\relax
\EndOfBibitem
\bibitem[Runge and Gross(1984)Runge, and Gross]{runge1984density}
Runge,~E.; Gross,~E.~K. Density-functional theory for time-dependent systems. \emph{Physical review letters} \textbf{1984}, \emph{52}, 997\relax
\mciteBstWouldAddEndPuncttrue
\mciteSetBstMidEndSepPunct{\mcitedefaultmidpunct}
{\mcitedefaultendpunct}{\mcitedefaultseppunct}\relax
\EndOfBibitem
\bibitem[Burke \latin{et~al.}(2005)Burke, Werschnik, and Gross]{burke2005time}
Burke,~K.; Werschnik,~J.; Gross,~E. Time-dependent density functional theory: Past, present, and future. \emph{The Journal of chemical physics} \textbf{2005}, \emph{123}, 062206\relax
\mciteBstWouldAddEndPuncttrue
\mciteSetBstMidEndSepPunct{\mcitedefaultmidpunct}
{\mcitedefaultendpunct}{\mcitedefaultseppunct}\relax
\EndOfBibitem
\bibitem[Casida and Huix-Rotllant(2012)Casida, and Huix-Rotllant]{casida2012progress}
Casida,~M.~E.; Huix-Rotllant,~M. Progress in time-dependent density-functional theory. \emph{Annual review of physical chemistry} \textbf{2012}, \emph{63}, 287--323\relax
\mciteBstWouldAddEndPuncttrue
\mciteSetBstMidEndSepPunct{\mcitedefaultmidpunct}
{\mcitedefaultendpunct}{\mcitedefaultseppunct}\relax
\EndOfBibitem
\bibitem[Rowe(1968)]{rowe1968equations}
Rowe,~D. Equations-of-motion method and the extended shell model. \emph{Reviews of Modern Physics} \textbf{1968}, \emph{40}, 153\relax
\mciteBstWouldAddEndPuncttrue
\mciteSetBstMidEndSepPunct{\mcitedefaultmidpunct}
{\mcitedefaultendpunct}{\mcitedefaultseppunct}\relax
\EndOfBibitem
\bibitem[Stanton and Bartlett(1993)Stanton, and Bartlett]{stanton1993equation}
Stanton,~J.~F.; Bartlett,~R.~J. The equation of motion coupled-cluster method. A systematic biorthogonal approach to molecular excitation energies, transition probabilities, and excited state properties. \emph{The Journal of chemical physics} \textbf{1993}, \emph{98}, 7029--7039\relax
\mciteBstWouldAddEndPuncttrue
\mciteSetBstMidEndSepPunct{\mcitedefaultmidpunct}
{\mcitedefaultendpunct}{\mcitedefaultseppunct}\relax
\EndOfBibitem
\bibitem[Krylov(2008)]{krylov2008equation}
Krylov,~A.~I. Equation-of-motion coupled-cluster methods for open-shell and electronically excited species: The hitchhiker's guide to Fock space. \emph{Annu. Rev. Phys. Chem.} \textbf{2008}, \emph{59}, 433--462\relax
\mciteBstWouldAddEndPuncttrue
\mciteSetBstMidEndSepPunct{\mcitedefaultmidpunct}
{\mcitedefaultendpunct}{\mcitedefaultseppunct}\relax
\EndOfBibitem
\bibitem[Monkhorst(1977)]{monkhorst1977calculation}
Monkhorst,~H.~J. Calculation of properties with the coupled-cluster method. \emph{International Journal of Quantum Chemistry} \textbf{1977}, \emph{12}, 421--432\relax
\mciteBstWouldAddEndPuncttrue
\mciteSetBstMidEndSepPunct{\mcitedefaultmidpunct}
{\mcitedefaultendpunct}{\mcitedefaultseppunct}\relax
\EndOfBibitem
\bibitem[Dalgaard and Monkhorst(1983)Dalgaard, and Monkhorst]{dalgaard1983some}
Dalgaard,~E.; Monkhorst,~H.~J. Some aspects of the time-dependent coupled-cluster approach to dynamic response functions. \emph{Physical Review A} \textbf{1983}, \emph{28}, 1217\relax
\mciteBstWouldAddEndPuncttrue
\mciteSetBstMidEndSepPunct{\mcitedefaultmidpunct}
{\mcitedefaultendpunct}{\mcitedefaultseppunct}\relax
\EndOfBibitem
\bibitem[Sekino and Bartlett(1984)Sekino, and Bartlett]{sekino1984linear}
Sekino,~H.; Bartlett,~R.~J. A linear response, coupled-cluster theory for excitation energy. \emph{International Journal of Quantum Chemistry} \textbf{1984}, \emph{26}, 255--265\relax
\mciteBstWouldAddEndPuncttrue
\mciteSetBstMidEndSepPunct{\mcitedefaultmidpunct}
{\mcitedefaultendpunct}{\mcitedefaultseppunct}\relax
\EndOfBibitem
\bibitem[Koch \latin{et~al.}(1990)Koch, Jensen, Jo/rgensen, and Helgaker]{koch1990excitation}
Koch,~H.; Jensen,~H. J.~A.; Jo/rgensen,~P.; Helgaker,~T. Excitation energies from the coupled cluster singles and doubles linear response function (CCSDLR). Applications to Be, CH+, CO, and H2O. \emph{The Journal of chemical physics} \textbf{1990}, \emph{93}, 3345--3350\relax
\mciteBstWouldAddEndPuncttrue
\mciteSetBstMidEndSepPunct{\mcitedefaultmidpunct}
{\mcitedefaultendpunct}{\mcitedefaultseppunct}\relax
\EndOfBibitem
\bibitem[Koch and J{\o}rgensen(1990)Koch, and J{\o}rgensen]{koch1990coupled}
Koch,~H.; J{\o}rgensen,~P. Coupled cluster response functions. \emph{The Journal of chemical physics} \textbf{1990}, \emph{93}, 3333\relax
\mciteBstWouldAddEndPuncttrue
\mciteSetBstMidEndSepPunct{\mcitedefaultmidpunct}
{\mcitedefaultendpunct}{\mcitedefaultseppunct}\relax
\EndOfBibitem
\bibitem[Rico and Head-Gordon(1993)Rico, and Head-Gordon]{rico1993single}
Rico,~R.~J.; Head-Gordon,~M. Single-reference theories of molecular excited states with single and double substitutions. \emph{Chemical physics letters} \textbf{1993}, \emph{213}, 224--232\relax
\mciteBstWouldAddEndPuncttrue
\mciteSetBstMidEndSepPunct{\mcitedefaultmidpunct}
{\mcitedefaultendpunct}{\mcitedefaultseppunct}\relax
\EndOfBibitem
\bibitem[Koch \latin{et~al.}(1994)Koch, Kobayashi, Sanchez~de Mer{\'a}s, and J{\o}rgensen]{koch1994calculation}
Koch,~H.; Kobayashi,~R.; Sanchez~de Mer{\'a}s,~A.; J{\o}rgensen,~P. Calculation of size-intensive transition moments from the coupled cluster singles and doubles linear response function. \emph{The Journal of chemical physics} \textbf{1994}, \emph{100}, 4393--4400\relax
\mciteBstWouldAddEndPuncttrue
\mciteSetBstMidEndSepPunct{\mcitedefaultmidpunct}
{\mcitedefaultendpunct}{\mcitedefaultseppunct}\relax
\EndOfBibitem
\bibitem[Sneskov and Christiansen(2012)Sneskov, and Christiansen]{sneskov2012excited}
Sneskov,~K.; Christiansen,~O. Excited state coupled cluster methods. \emph{Wiley Interdisciplinary Reviews: Computational Molecular Science} \textbf{2012}, \emph{2}, 566--584\relax
\mciteBstWouldAddEndPuncttrue
\mciteSetBstMidEndSepPunct{\mcitedefaultmidpunct}
{\mcitedefaultendpunct}{\mcitedefaultseppunct}\relax
\EndOfBibitem
\bibitem[Tozer and Handy(1998)Tozer, and Handy]{tozer1998improving}
Tozer,~D.~J.; Handy,~N.~C. Improving virtual Kohn--Sham orbitals and eigenvalues: Application to excitation energies and static polarizabilities. \emph{The Journal of chemical physics} \textbf{1998}, \emph{109}, 10180--10189\relax
\mciteBstWouldAddEndPuncttrue
\mciteSetBstMidEndSepPunct{\mcitedefaultmidpunct}
{\mcitedefaultendpunct}{\mcitedefaultseppunct}\relax
\EndOfBibitem
\bibitem[Casida \latin{et~al.}(1998)Casida, Jamorski, Casida, and Salahub]{casida1998molecular}
Casida,~M.~E.; Jamorski,~C.; Casida,~K.~C.; Salahub,~D.~R. Molecular excitation energies to high-lying bound states from time-dependent density-functional response theory: Characterization and correction of the time-dependent local density approximation ionization threshold. \emph{The Journal of chemical physics} \textbf{1998}, \emph{108}, 4439--4449\relax
\mciteBstWouldAddEndPuncttrue
\mciteSetBstMidEndSepPunct{\mcitedefaultmidpunct}
{\mcitedefaultendpunct}{\mcitedefaultseppunct}\relax
\EndOfBibitem
\bibitem[Casida and Salahub(2000)Casida, and Salahub]{casida2000asymptotic}
Casida,~M.~E.; Salahub,~D.~R. Asymptotic correction approach to improving approximate exchange--correlation potentials: Time-dependent density-functional theory calculations of molecular excitation spectra. \emph{The Journal of Chemical Physics} \textbf{2000}, \emph{113}, 8918--8935\relax
\mciteBstWouldAddEndPuncttrue
\mciteSetBstMidEndSepPunct{\mcitedefaultmidpunct}
{\mcitedefaultendpunct}{\mcitedefaultseppunct}\relax
\EndOfBibitem
\bibitem[Tozer and Handy(2003)Tozer, and Handy]{tozer2003importance}
Tozer,~D.~J.; Handy,~N.~C. The importance of the asymptotic exchange-correlation potential in density functional theory. \emph{Molecular Physics} \textbf{2003}, \emph{101}, 2669--2675\relax
\mciteBstWouldAddEndPuncttrue
\mciteSetBstMidEndSepPunct{\mcitedefaultmidpunct}
{\mcitedefaultendpunct}{\mcitedefaultseppunct}\relax
\EndOfBibitem
\bibitem[Sobolewski and Domcke(2003)Sobolewski, and Domcke]{sobolewski2003ab}
Sobolewski,~A.~L.; Domcke,~W. Ab initio study of the excited-state coupled electron--proton-transfer process in the 2-aminopyridine dimer. \emph{Chemical Physics} \textbf{2003}, \emph{294}, 73--83\relax
\mciteBstWouldAddEndPuncttrue
\mciteSetBstMidEndSepPunct{\mcitedefaultmidpunct}
{\mcitedefaultendpunct}{\mcitedefaultseppunct}\relax
\EndOfBibitem
\bibitem[Dreuw \latin{et~al.}(2003)Dreuw, Weisman, and Head-Gordon]{dreuw2003long}
Dreuw,~A.; Weisman,~J.~L.; Head-Gordon,~M. Long-range charge-transfer excited states in time-dependent density functional theory require non-local exchange. \emph{The Journal of chemical physics} \textbf{2003}, \emph{119}, 2943--2946\relax
\mciteBstWouldAddEndPuncttrue
\mciteSetBstMidEndSepPunct{\mcitedefaultmidpunct}
{\mcitedefaultendpunct}{\mcitedefaultseppunct}\relax
\EndOfBibitem
\bibitem[Dreuw and Head-Gordon(2004)Dreuw, and Head-Gordon]{dreuw2004failure}
Dreuw,~A.; Head-Gordon,~M. Failure of time-dependent density functional theory for long-range charge-transfer excited states: the zincbacteriochlorin- bacteriochlorin and bacteriochlorophyll- spheroidene complexes. \emph{Journal of the American Chemical Society} \textbf{2004}, \emph{126}, 4007--4016\relax
\mciteBstWouldAddEndPuncttrue
\mciteSetBstMidEndSepPunct{\mcitedefaultmidpunct}
{\mcitedefaultendpunct}{\mcitedefaultseppunct}\relax
\EndOfBibitem
\bibitem[Mester and K{\'a}llay(2022)Mester, and K{\'a}llay]{mester2022charge}
Mester,~D.; K{\'a}llay,~M. Charge-transfer excitations within density functional theory: how accurate are the most recommended approaches? \emph{Journal of Chemical Theory and Computation} \textbf{2022}, \emph{18}, 1646--1662\relax
\mciteBstWouldAddEndPuncttrue
\mciteSetBstMidEndSepPunct{\mcitedefaultmidpunct}
{\mcitedefaultendpunct}{\mcitedefaultseppunct}\relax
\EndOfBibitem
\bibitem[Kozma \latin{et~al.}(2020)Kozma, Tajti, Demoulin, Izs{\'a}k, Nooijen, and Szalay]{kozma2020new}
Kozma,~B.; Tajti,~A.; Demoulin,~B.; Izs{\'a}k,~R.; Nooijen,~M.; Szalay,~P.~G. A new benchmark set for excitation energy of charge transfer states: systematic investigation of coupled cluster type methods. \emph{Journal of Chemical Theory and Computation} \textbf{2020}, \emph{16}, 4213--4225\relax
\mciteBstWouldAddEndPuncttrue
\mciteSetBstMidEndSepPunct{\mcitedefaultmidpunct}
{\mcitedefaultendpunct}{\mcitedefaultseppunct}\relax
\EndOfBibitem
\bibitem[Izs{\'a}k(2020)]{izsak2020single}
Izs{\'a}k,~R. Single-reference coupled cluster methods for computing excitation energies in large molecules: The efficiency and accuracy of approximations. \emph{Wiley Interdisciplinary Reviews: Computational Molecular Science} \textbf{2020}, \emph{10}, e1445\relax
\mciteBstWouldAddEndPuncttrue
\mciteSetBstMidEndSepPunct{\mcitedefaultmidpunct}
{\mcitedefaultendpunct}{\mcitedefaultseppunct}\relax
\EndOfBibitem
\bibitem[Ziegler \latin{et~al.}(1977)Ziegler, Rauk, and Baerends]{ziegler1977calculation}
Ziegler,~T.; Rauk,~A.; Baerends,~E.~J. On the calculation of multiplet energies by the Hartree-Fock-Slater method. \emph{Theoretica chimica acta} \textbf{1977}, \emph{43}, 261--271\relax
\mciteBstWouldAddEndPuncttrue
\mciteSetBstMidEndSepPunct{\mcitedefaultmidpunct}
{\mcitedefaultendpunct}{\mcitedefaultseppunct}\relax
\EndOfBibitem
\bibitem[Kowalczyk \latin{et~al.}(2011)Kowalczyk, Yost, and Voorhis]{kowalczyk2011assessment}
Kowalczyk,~T.; Yost,~S.~R.; Voorhis,~T.~V. Assessment of the $\Delta$SCF density functional theory approach for electronic excitations in organic dyes. \emph{The Journal of chemical physics} \textbf{2011}, \emph{134}\relax
\mciteBstWouldAddEndPuncttrue
\mciteSetBstMidEndSepPunct{\mcitedefaultmidpunct}
{\mcitedefaultendpunct}{\mcitedefaultseppunct}\relax
\EndOfBibitem
\bibitem[Gilbert \latin{et~al.}(2008)Gilbert, Besley, and Gill]{gilbert2008self}
Gilbert,~A.~T.; Besley,~N.~A.; Gill,~P.~M. Self-consistent field calculations of excited states using the maximum overlap method (MOM). \emph{The Journal of Physical Chemistry A} \textbf{2008}, \emph{112}, 13164--13171\relax
\mciteBstWouldAddEndPuncttrue
\mciteSetBstMidEndSepPunct{\mcitedefaultmidpunct}
{\mcitedefaultendpunct}{\mcitedefaultseppunct}\relax
\EndOfBibitem
\bibitem[Besley \latin{et~al.}(2009)Besley, Gilbert, and Gill]{besley2009self}
Besley,~N.~A.; Gilbert,~A.~T.; Gill,~P.~M. Self-consistent-field calculations of core excited states. \emph{The Journal of chemical physics} \textbf{2009}, \emph{130}, 124308\relax
\mciteBstWouldAddEndPuncttrue
\mciteSetBstMidEndSepPunct{\mcitedefaultmidpunct}
{\mcitedefaultendpunct}{\mcitedefaultseppunct}\relax
\EndOfBibitem
\bibitem[Barca \latin{et~al.}(2018)Barca, Gilbert, and Gill]{barca2018simple}
Barca,~G.~M.; Gilbert,~A.~T.; Gill,~P.~M. Simple models for difficult electronic excitations. \emph{Journal of chemical theory and computation} \textbf{2018}, \emph{14}, 1501--1509\relax
\mciteBstWouldAddEndPuncttrue
\mciteSetBstMidEndSepPunct{\mcitedefaultmidpunct}
{\mcitedefaultendpunct}{\mcitedefaultseppunct}\relax
\EndOfBibitem
\bibitem[Carter-Fenk and Herbert(2020)Carter-Fenk, and Herbert]{carter2020state}
Carter-Fenk,~K.; Herbert,~J.~M. State-targeted energy projection: A simple and robust approach to orbital relaxation of non-Aufbau self-consistent field solutions. \emph{Journal of Chemical Theory and Computation} \textbf{2020}, \emph{16}, 5067--5082\relax
\mciteBstWouldAddEndPuncttrue
\mciteSetBstMidEndSepPunct{\mcitedefaultmidpunct}
{\mcitedefaultendpunct}{\mcitedefaultseppunct}\relax
\EndOfBibitem
\bibitem[Burton and Wales(2020)Burton, and Wales]{burton2020energy}
Burton,~H.~G.; Wales,~D.~J. Energy landscapes for electronic structure. \emph{Journal of Chemical Theory and Computation} \textbf{2020}, \emph{17}, 151--169\relax
\mciteBstWouldAddEndPuncttrue
\mciteSetBstMidEndSepPunct{\mcitedefaultmidpunct}
{\mcitedefaultendpunct}{\mcitedefaultseppunct}\relax
\EndOfBibitem
\bibitem[Dreuw and Head-Gordon(2005)Dreuw, and Head-Gordon]{dreuw2005single}
Dreuw,~A.; Head-Gordon,~M. Single-reference ab initio methods for the calculation of excited states of large molecules. \emph{Chemical reviews} \textbf{2005}, \emph{105}, 4009--4037\relax
\mciteBstWouldAddEndPuncttrue
\mciteSetBstMidEndSepPunct{\mcitedefaultmidpunct}
{\mcitedefaultendpunct}{\mcitedefaultseppunct}\relax
\EndOfBibitem
\bibitem[Liu \latin{et~al.}(2012)Liu, Fatehi, Shao, Veldkamp, and Subotnik]{liu2012communication}
Liu,~X.; Fatehi,~S.; Shao,~Y.; Veldkamp,~B.~S.; Subotnik,~J.~E. Communication: Adjusting charge transfer state energies for configuration interaction singles: Without any parameterization and with minimal cost. \emph{The Journal of chemical physics} \textbf{2012}, \emph{136}, 161101\relax
\mciteBstWouldAddEndPuncttrue
\mciteSetBstMidEndSepPunct{\mcitedefaultmidpunct}
{\mcitedefaultendpunct}{\mcitedefaultseppunct}\relax
\EndOfBibitem
\bibitem[Liu and Subotnik(2014)Liu, and Subotnik]{liu2014variationally}
Liu,~X.; Subotnik,~J.~E. The variationally orbital-adapted configuration interaction singles (VOA-CIS) approach to electronically excited states. \emph{Journal of Chemical Theory and Computation} \textbf{2014}, \emph{10}, 1004--1020\relax
\mciteBstWouldAddEndPuncttrue
\mciteSetBstMidEndSepPunct{\mcitedefaultmidpunct}
{\mcitedefaultendpunct}{\mcitedefaultseppunct}\relax
\EndOfBibitem
\bibitem[Shea and Neuscamman(2018)Shea, and Neuscamman]{shea2018communication}
Shea,~J.~A.; Neuscamman,~E. Communication: A mean field platform for excited state quantum chemistry. \emph{The Journal of chemical physics} \textbf{2018}, \emph{149}, 081101\relax
\mciteBstWouldAddEndPuncttrue
\mciteSetBstMidEndSepPunct{\mcitedefaultmidpunct}
{\mcitedefaultendpunct}{\mcitedefaultseppunct}\relax
\EndOfBibitem
\bibitem[Shea \latin{et~al.}(2020)Shea, Gwin, and Neuscamman]{shea2020generalized}
Shea,~J.~A.; Gwin,~E.; Neuscamman,~E. A generalized variational principle with applications to excited state mean field theory. \emph{Journal of chemical theory and computation} \textbf{2020}, \emph{16}, 1526--1540\relax
\mciteBstWouldAddEndPuncttrue
\mciteSetBstMidEndSepPunct{\mcitedefaultmidpunct}
{\mcitedefaultendpunct}{\mcitedefaultseppunct}\relax
\EndOfBibitem
\bibitem[Hardikar and Neuscamman(2020)Hardikar, and Neuscamman]{hardikar2020self}
Hardikar,~T.~S.; Neuscamman,~E. A self-consistent field formulation of excited state mean field theory. \emph{The Journal of chemical physics} \textbf{2020}, \emph{153}, 164108\relax
\mciteBstWouldAddEndPuncttrue
\mciteSetBstMidEndSepPunct{\mcitedefaultmidpunct}
{\mcitedefaultendpunct}{\mcitedefaultseppunct}\relax
\EndOfBibitem
\bibitem[Kossoski and Loos(2022)Kossoski, and Loos]{kossoski2022state}
Kossoski,~F.; Loos,~P.-F. State-Specific Configuration Interaction for Excited States. \emph{arXiv preprint arXiv:2211.03048} \textbf{2022}, \relax
\mciteBstWouldAddEndPunctfalse
\mciteSetBstMidEndSepPunct{\mcitedefaultmidpunct}
{}{\mcitedefaultseppunct}\relax
\EndOfBibitem
\bibitem[Kossoski and Loos(2023)Kossoski, and Loos]{kossoski2023seniority}
Kossoski,~F.; Loos,~P.-F. Seniority and Hierarchy Configuration Interaction for Radicals and Excited States. \emph{arXiv preprint arXiv:2308.14618} \textbf{2023}, \relax
\mciteBstWouldAddEndPunctfalse
\mciteSetBstMidEndSepPunct{\mcitedefaultmidpunct}
{}{\mcitedefaultseppunct}\relax
\EndOfBibitem
\bibitem[Burton(2022)]{burton2022energy}
Burton,~H.~G. Energy Landscape of State-Specific Electronic Structure Theory. \emph{Journal of Chemical Theory and Computation} \textbf{2022}, \emph{18}, 1512--1526\relax
\mciteBstWouldAddEndPuncttrue
\mciteSetBstMidEndSepPunct{\mcitedefaultmidpunct}
{\mcitedefaultendpunct}{\mcitedefaultseppunct}\relax
\EndOfBibitem
\bibitem[Clune \latin{et~al.}(2020)Clune, Shea, and Neuscamman]{clune2020n5}
Clune,~R.; Shea,~J.~A.; Neuscamman,~E. N5-scaling excited-state-specific perturbation theory. \emph{Journal of chemical theory and computation} \textbf{2020}, \emph{16}, 6132--6141\relax
\mciteBstWouldAddEndPuncttrue
\mciteSetBstMidEndSepPunct{\mcitedefaultmidpunct}
{\mcitedefaultendpunct}{\mcitedefaultseppunct}\relax
\EndOfBibitem
\bibitem[Clune \latin{et~al.}(2023)Clune, Shea, Hardikar, Tuckman, and Neuscamman]{clune2023studying}
Clune,~R.; Shea,~J.~A.; Hardikar,~T.~S.; Tuckman,~H.; Neuscamman,~E. Studying excited-state-specific perturbation theory on the Thiel set. \emph{The Journal of Chemical Physics} \textbf{2023}, \emph{158}\relax
\mciteBstWouldAddEndPuncttrue
\mciteSetBstMidEndSepPunct{\mcitedefaultmidpunct}
{\mcitedefaultendpunct}{\mcitedefaultseppunct}\relax
\EndOfBibitem
\bibitem[Mayhall and Raghavachari(2010)Mayhall, and Raghavachari]{mayhall2010multiple}
Mayhall,~N.~J.; Raghavachari,~K. Multiple solutions to the single-reference CCSD equations for NiH. \emph{Journal of Chemical Theory and Computation} \textbf{2010}, \emph{6}, 2714--2720\relax
\mciteBstWouldAddEndPuncttrue
\mciteSetBstMidEndSepPunct{\mcitedefaultmidpunct}
{\mcitedefaultendpunct}{\mcitedefaultseppunct}\relax
\EndOfBibitem
\bibitem[Lee \latin{et~al.}(2019)Lee, Small, and Head-Gordon]{lee2019excited}
Lee,~J.; Small,~D.~W.; Head-Gordon,~M. Excited states via coupled cluster theory without equation-of-motion methods: Seeking higher roots with application to doubly excited states and double core hole states. \emph{The Journal of chemical physics} \textbf{2019}, \emph{151}, 214103\relax
\mciteBstWouldAddEndPuncttrue
\mciteSetBstMidEndSepPunct{\mcitedefaultmidpunct}
{\mcitedefaultendpunct}{\mcitedefaultseppunct}\relax
\EndOfBibitem
\bibitem[Kossoski \latin{et~al.}(2021)Kossoski, Marie, Scemama, Caffarel, and Loos]{kossoski2021excited}
Kossoski,~F.; Marie,~A.; Scemama,~A.; Caffarel,~M.; Loos,~P.-F. Excited States from State-Specific Orbital-Optimized Pair Coupled Cluster. \emph{Journal of Chemical Theory and Computation} \textbf{2021}, \emph{17}, 4756--4768\relax
\mciteBstWouldAddEndPuncttrue
\mciteSetBstMidEndSepPunct{\mcitedefaultmidpunct}
{\mcitedefaultendpunct}{\mcitedefaultseppunct}\relax
\EndOfBibitem
\bibitem[Marie \latin{et~al.}(2021)Marie, Kossoski, and Loos]{marie2021variational}
Marie,~A.; Kossoski,~F.; Loos,~P.-F. Variational coupled cluster for ground and excited states. \emph{The Journal of Chemical Physics} \textbf{2021}, \emph{155}\relax
\mciteBstWouldAddEndPuncttrue
\mciteSetBstMidEndSepPunct{\mcitedefaultmidpunct}
{\mcitedefaultendpunct}{\mcitedefaultseppunct}\relax
\EndOfBibitem
\bibitem[Rishi \latin{et~al.}(2023)Rishi, Ravi, Perera, and Bartlett]{rishi2023dark}
Rishi,~V.; Ravi,~M.; Perera,~A.; Bartlett,~R.~J. Dark doubly excited states with modified coupled cluster models: A reliable compromise between cost and accuracy? \emph{The Journal of Physical Chemistry A} \textbf{2023}, \emph{127}, 828--834\relax
\mciteBstWouldAddEndPuncttrue
\mciteSetBstMidEndSepPunct{\mcitedefaultmidpunct}
{\mcitedefaultendpunct}{\mcitedefaultseppunct}\relax
\EndOfBibitem
\bibitem[Knowles and Werner(1985)Knowles, and Werner]{knowles1985efficient}
Knowles,~P.~J.; Werner,~H.-J. An efficient second-order MC SCF method for long configuration expansions. \emph{Chemical physics letters} \textbf{1985}, \emph{115}, 259--267\relax
\mciteBstWouldAddEndPuncttrue
\mciteSetBstMidEndSepPunct{\mcitedefaultmidpunct}
{\mcitedefaultendpunct}{\mcitedefaultseppunct}\relax
\EndOfBibitem
\bibitem[Werner and Knowles(1985)Werner, and Knowles]{werner1985second}
Werner,~H.-J.; Knowles,~P.~J. A second order multiconfiguration SCF procedure with optimum convergence. \emph{The Journal of chemical physics} \textbf{1985}, \emph{82}, 5053--5063\relax
\mciteBstWouldAddEndPuncttrue
\mciteSetBstMidEndSepPunct{\mcitedefaultmidpunct}
{\mcitedefaultendpunct}{\mcitedefaultseppunct}\relax
\EndOfBibitem
\bibitem[Ruedenberg \latin{et~al.}(1982)Ruedenberg, Schmidt, Gilbert, and Elbert]{ruedenberg1982atoms}
Ruedenberg,~K.; Schmidt,~M.~W.; Gilbert,~M.~M.; Elbert,~S. Are atoms intrinsic to molecular electronic wavefunctions? I. The FORS model. \emph{Chemical Physics} \textbf{1982}, \emph{71}, 41--49\relax
\mciteBstWouldAddEndPuncttrue
\mciteSetBstMidEndSepPunct{\mcitedefaultmidpunct}
{\mcitedefaultendpunct}{\mcitedefaultseppunct}\relax
\EndOfBibitem
\bibitem[Roos(1987)]{roos1987complete}
Roos,~B.~O. The complete active space self-consistent field method and its applications in electronic structure calculations. \emph{Advances in Chemical Physics: Ab Initio Methods in Quantum Chemistry Part 2} \textbf{1987}, \emph{69}, 399--445\relax
\mciteBstWouldAddEndPuncttrue
\mciteSetBstMidEndSepPunct{\mcitedefaultmidpunct}
{\mcitedefaultendpunct}{\mcitedefaultseppunct}\relax
\EndOfBibitem
\bibitem[Tran \latin{et~al.}(2019)Tran, Shea, and Neuscamman]{tran2019tracking}
Tran,~L.~N.; Shea,~J.~A.; Neuscamman,~E. Tracking excited states in wave function optimization using density matrices and variational principles. \emph{Journal of chemical theory and computation} \textbf{2019}, \emph{15}, 4790--4803\relax
\mciteBstWouldAddEndPuncttrue
\mciteSetBstMidEndSepPunct{\mcitedefaultmidpunct}
{\mcitedefaultendpunct}{\mcitedefaultseppunct}\relax
\EndOfBibitem
\bibitem[Tran and Neuscamman(2020)Tran, and Neuscamman]{tran2020improving}
Tran,~L.~N.; Neuscamman,~E. Improving excited-state potential energy surfaces via optimal orbital shapes. \emph{The Journal of Physical Chemistry A} \textbf{2020}, \emph{124}, 8273--8279\relax
\mciteBstWouldAddEndPuncttrue
\mciteSetBstMidEndSepPunct{\mcitedefaultmidpunct}
{\mcitedefaultendpunct}{\mcitedefaultseppunct}\relax
\EndOfBibitem
\bibitem[Hanscam and Neuscamman(2022)Hanscam, and Neuscamman]{hanscam2022applying}
Hanscam,~R.; Neuscamman,~E. Applying generalized variational principles to excited-state-specific complete active space self-consistent field theory. \emph{Journal of Chemical Theory and Computation} \textbf{2022}, \emph{18}, 6608--6621\relax
\mciteBstWouldAddEndPuncttrue
\mciteSetBstMidEndSepPunct{\mcitedefaultmidpunct}
{\mcitedefaultendpunct}{\mcitedefaultseppunct}\relax
\EndOfBibitem
\bibitem[Kowalczyk \latin{et~al.}(2013)Kowalczyk, Tsuchimochi, Chen, Top, and Van~Voorhis]{kowalczyk2013excitation}
Kowalczyk,~T.; Tsuchimochi,~T.; Chen,~P.-T.; Top,~L.; Van~Voorhis,~T. Excitation energies and Stokes shifts from a restricted open-shell Kohn-Sham approach. \emph{The Journal of chemical physics} \textbf{2013}, \emph{138}, 164101\relax
\mciteBstWouldAddEndPuncttrue
\mciteSetBstMidEndSepPunct{\mcitedefaultmidpunct}
{\mcitedefaultendpunct}{\mcitedefaultseppunct}\relax
\EndOfBibitem
\bibitem[Hait and Head-Gordon(2020)Hait, and Head-Gordon]{hait2020excited}
Hait,~D.; Head-Gordon,~M. Excited state orbital optimization via minimizing the square of the gradient: General approach and application to singly and doubly excited states via density functional theory. \emph{Journal of chemical theory and computation} \textbf{2020}, \emph{16}, 1699--1710\relax
\mciteBstWouldAddEndPuncttrue
\mciteSetBstMidEndSepPunct{\mcitedefaultmidpunct}
{\mcitedefaultendpunct}{\mcitedefaultseppunct}\relax
\EndOfBibitem
\bibitem[Hait and Head-Gordon(2021)Hait, and Head-Gordon]{hait2021orbital}
Hait,~D.; Head-Gordon,~M. Orbital optimized density functional theory for electronic excited states. \emph{The journal of physical chemistry letters} \textbf{2021}, \emph{12}, 4517--4529\relax
\mciteBstWouldAddEndPuncttrue
\mciteSetBstMidEndSepPunct{\mcitedefaultmidpunct}
{\mcitedefaultendpunct}{\mcitedefaultseppunct}\relax
\EndOfBibitem
\bibitem[Zhao and Neuscamman(2019)Zhao, and Neuscamman]{zhao2019density}
Zhao,~L.; Neuscamman,~E. Density functional extension to excited-state mean-field theory. \emph{Journal of chemical theory and computation} \textbf{2019}, \emph{16}, 164--178\relax
\mciteBstWouldAddEndPuncttrue
\mciteSetBstMidEndSepPunct{\mcitedefaultmidpunct}
{\mcitedefaultendpunct}{\mcitedefaultseppunct}\relax
\EndOfBibitem
\bibitem[Levi \latin{et~al.}(2020)Levi, Ivanov, and J{\'o}nsson]{levi2020variational}
Levi,~G.; Ivanov,~A.~V.; J{\'o}nsson,~H. Variational density functional calculations of excited states via direct optimization. \emph{Journal of Chemical Theory and Computation} \textbf{2020}, \emph{16}, 6968--6982\relax
\mciteBstWouldAddEndPuncttrue
\mciteSetBstMidEndSepPunct{\mcitedefaultmidpunct}
{\mcitedefaultendpunct}{\mcitedefaultseppunct}\relax
\EndOfBibitem
\bibitem[Riplinger and Neese(2013)Riplinger, and Neese]{riplinger2013efficient}
Riplinger,~C.; Neese,~F. An efficient and near linear scaling pair natural orbital based local coupled cluster method. \emph{The Journal of chemical physics} \textbf{2013}, \emph{138}, 034106\relax
\mciteBstWouldAddEndPuncttrue
\mciteSetBstMidEndSepPunct{\mcitedefaultmidpunct}
{\mcitedefaultendpunct}{\mcitedefaultseppunct}\relax
\EndOfBibitem
\bibitem[Riplinger \latin{et~al.}(2013)Riplinger, Sandhoefer, Hansen, and Neese]{riplinger2013natural}
Riplinger,~C.; Sandhoefer,~B.; Hansen,~A.; Neese,~F. Natural triple excitations in local coupled cluster calculations with pair natural orbitals. \emph{The Journal of chemical physics} \textbf{2013}, \emph{139}, 134101\relax
\mciteBstWouldAddEndPuncttrue
\mciteSetBstMidEndSepPunct{\mcitedefaultmidpunct}
{\mcitedefaultendpunct}{\mcitedefaultseppunct}\relax
\EndOfBibitem
\bibitem[Riplinger \latin{et~al.}(2016)Riplinger, Pinski, Becker, Valeev, and Neese]{riplinger2016sparse}
Riplinger,~C.; Pinski,~P.; Becker,~U.; Valeev,~E.~F.; Neese,~F. Sparse maps—A systematic infrastructure for reduced-scaling electronic structure methods. II. Linear scaling domain based pair natural orbital coupled cluster theory. \emph{The Journal of chemical physics} \textbf{2016}, \emph{144}, 024109\relax
\mciteBstWouldAddEndPuncttrue
\mciteSetBstMidEndSepPunct{\mcitedefaultmidpunct}
{\mcitedefaultendpunct}{\mcitedefaultseppunct}\relax
\EndOfBibitem
\bibitem[Saitow \latin{et~al.}(2017)Saitow, Becker, Riplinger, Valeev, and Neese]{saitow2017new}
Saitow,~M.; Becker,~U.; Riplinger,~C.; Valeev,~E.~F.; Neese,~F. A new near-linear scaling, efficient and accurate, open-shell domain-based local pair natural orbital coupled cluster singles and doubles theory. \emph{The Journal of chemical physics} \textbf{2017}, \emph{146}, 164105\relax
\mciteBstWouldAddEndPuncttrue
\mciteSetBstMidEndSepPunct{\mcitedefaultmidpunct}
{\mcitedefaultendpunct}{\mcitedefaultseppunct}\relax
\EndOfBibitem
\bibitem[Guo \latin{et~al.}(2018)Guo, Riplinger, Becker, Liakos, Minenkov, Cavallo, and Neese]{guo2018communication}
Guo,~Y.; Riplinger,~C.; Becker,~U.; Liakos,~D.~G.; Minenkov,~Y.; Cavallo,~L.; Neese,~F. Communication: An improved linear scaling perturbative triples correction for the domain based local pair-natural orbital based singles and doubles coupled cluster method [DLPNO-CCSD (T)]. \emph{The Journal of chemical physics} \textbf{2018}, \emph{148}, 011101\relax
\mciteBstWouldAddEndPuncttrue
\mciteSetBstMidEndSepPunct{\mcitedefaultmidpunct}
{\mcitedefaultendpunct}{\mcitedefaultseppunct}\relax
\EndOfBibitem
\bibitem[Frank and H{\"a}ttig(2018)Frank, and H{\"a}ttig]{frank2018pair}
Frank,~M.~S.; H{\"a}ttig,~C. A pair natural orbital based implementation of CCSD excitation energies within the framework of linear response theory. \emph{The Journal of Chemical Physics} \textbf{2018}, \emph{148}\relax
\mciteBstWouldAddEndPuncttrue
\mciteSetBstMidEndSepPunct{\mcitedefaultmidpunct}
{\mcitedefaultendpunct}{\mcitedefaultseppunct}\relax
\EndOfBibitem
\bibitem[Helmich and H{\"a}ttig(2011)Helmich, and H{\"a}ttig]{helmich2011local}
Helmich,~B.; H{\"a}ttig,~C. Local pair natural orbitals for excited states. \emph{The Journal of chemical physics} \textbf{2011}, \emph{135}\relax
\mciteBstWouldAddEndPuncttrue
\mciteSetBstMidEndSepPunct{\mcitedefaultmidpunct}
{\mcitedefaultendpunct}{\mcitedefaultseppunct}\relax
\EndOfBibitem
\bibitem[Helmich and Haettig(2013)Helmich, and Haettig]{helmich2013pair}
Helmich,~B.; Haettig,~C. A pair natural orbital implementation of the coupled cluster model CC2 for excitation energies. \emph{The Journal of Chemical Physics} \textbf{2013}, \emph{139}\relax
\mciteBstWouldAddEndPuncttrue
\mciteSetBstMidEndSepPunct{\mcitedefaultmidpunct}
{\mcitedefaultendpunct}{\mcitedefaultseppunct}\relax
\EndOfBibitem
\bibitem[Cooper and Knowles(2010)Cooper, and Knowles]{cooper2010benchmark}
Cooper,~B.; Knowles,~P.~J. Benchmark studies of variational, unitary and extended coupled cluster methods. \emph{The Journal of chemical physics} \textbf{2010}, \emph{133}, 234102\relax
\mciteBstWouldAddEndPuncttrue
\mciteSetBstMidEndSepPunct{\mcitedefaultmidpunct}
{\mcitedefaultendpunct}{\mcitedefaultseppunct}\relax
\EndOfBibitem
\bibitem[Arponen(1983)]{arponen1983variational}
Arponen,~J. Variational principles and linked-cluster exp S expansions for static and dynamic many-body problems. \emph{Annals of Physics} \textbf{1983}, \emph{151}, 311--382\relax
\mciteBstWouldAddEndPuncttrue
\mciteSetBstMidEndSepPunct{\mcitedefaultmidpunct}
{\mcitedefaultendpunct}{\mcitedefaultseppunct}\relax
\EndOfBibitem
\bibitem[Arponen \latin{et~al.}(1987)Arponen, Bishop, and Pajanne]{arponen1987extended}
Arponen,~J.; Bishop,~R.; Pajanne,~E. Extended coupled-cluster method. I. Generalized coherent bosonization as a mapping of quantum theory into classical Hamiltonian mechanics. \emph{Physical Review A} \textbf{1987}, \emph{36}, 2519\relax
\mciteBstWouldAddEndPuncttrue
\mciteSetBstMidEndSepPunct{\mcitedefaultmidpunct}
{\mcitedefaultendpunct}{\mcitedefaultseppunct}\relax
\EndOfBibitem
\bibitem[Piecuch and Bartlett(1999)Piecuch, and Bartlett]{piecuch1999eomxcc}
Piecuch,~P.; Bartlett,~R.~J. \emph{Advances in Quantum Chemistry}; Elsevier, 1999; Vol.~34; pp 295--380\relax
\mciteBstWouldAddEndPuncttrue
\mciteSetBstMidEndSepPunct{\mcitedefaultmidpunct}
{\mcitedefaultendpunct}{\mcitedefaultseppunct}\relax
\EndOfBibitem
\bibitem[Fan \latin{et~al.}(2005)Fan, Kowalski, and Piecuch*]{fan2005non}
Fan,~P.-D.; Kowalski,~K.; Piecuch*,~P. Non-iterative corrections to extended coupled-cluster energies employing the generalized method of moments of coupled-cluster equations. \emph{Molecular Physics} \textbf{2005}, \emph{103}, 2191--2213\relax
\mciteBstWouldAddEndPuncttrue
\mciteSetBstMidEndSepPunct{\mcitedefaultmidpunct}
{\mcitedefaultendpunct}{\mcitedefaultseppunct}\relax
\EndOfBibitem
\bibitem[Fan and Piecuch(2006)Fan, and Piecuch]{fan2006usefulness}
Fan,~P.-D.; Piecuch,~P. The usefulness of exponential wave function expansions employing one-and two-body cluster operators in electronic structure theory: The extended and generalized coupled-cluster methods. \emph{Advances in Quantum Chemistry} \textbf{2006}, \emph{51}, 1--57\relax
\mciteBstWouldAddEndPuncttrue
\mciteSetBstMidEndSepPunct{\mcitedefaultmidpunct}
{\mcitedefaultendpunct}{\mcitedefaultseppunct}\relax
\EndOfBibitem
\bibitem[Van~Voorhis and Head-Gordon(2000)Van~Voorhis, and Head-Gordon]{van2000quadratic}
Van~Voorhis,~T.; Head-Gordon,~M. The quadratic coupled cluster doubles model. \emph{Chemical Physics Letters} \textbf{2000}, \emph{330}, 585--594\relax
\mciteBstWouldAddEndPuncttrue
\mciteSetBstMidEndSepPunct{\mcitedefaultmidpunct}
{\mcitedefaultendpunct}{\mcitedefaultseppunct}\relax
\EndOfBibitem
\bibitem[Goldstone(1957)]{goldstone1957derivation}
Goldstone,~J. Derivation of the Brueckner many-body theory. \emph{Proceedings of the Royal Society of London. Series A. Mathematical and Physical Sciences} \textbf{1957}, \emph{239}, 267--279\relax
\mciteBstWouldAddEndPuncttrue
\mciteSetBstMidEndSepPunct{\mcitedefaultmidpunct}
{\mcitedefaultendpunct}{\mcitedefaultseppunct}\relax
\EndOfBibitem
\bibitem[Bartlett and Purvis(1978)Bartlett, and Purvis]{bartlett1978many}
Bartlett,~R.~J.; Purvis,~G.~D. Many-body perturbation theory, coupled-pair many-electron theory, and the importance of quadruple excitations for the correlation problem. \emph{International Journal of Quantum Chemistry} \textbf{1978}, \emph{14}, 561--581\relax
\mciteBstWouldAddEndPuncttrue
\mciteSetBstMidEndSepPunct{\mcitedefaultmidpunct}
{\mcitedefaultendpunct}{\mcitedefaultseppunct}\relax
\EndOfBibitem
\bibitem[Pulay(1980)]{pulay1980convergence}
Pulay,~P. Convergence acceleration of iterative sequences. The case of SCF iteration. \emph{Chemical Physics Letters} \textbf{1980}, \emph{73}, 393--398\relax
\mciteBstWouldAddEndPuncttrue
\mciteSetBstMidEndSepPunct{\mcitedefaultmidpunct}
{\mcitedefaultendpunct}{\mcitedefaultseppunct}\relax
\EndOfBibitem
\bibitem[Hohenberg and Kohn(1964)Hohenberg, and Kohn]{hohenberg1964inhomogeneous}
Hohenberg,~P.; Kohn,~W. Inhomogeneous electron gas. \emph{Physical review} \textbf{1964}, \emph{136}, B864\relax
\mciteBstWouldAddEndPuncttrue
\mciteSetBstMidEndSepPunct{\mcitedefaultmidpunct}
{\mcitedefaultendpunct}{\mcitedefaultseppunct}\relax
\EndOfBibitem
\bibitem[Kohn and Sham(1965)Kohn, and Sham]{kohn1965self}
Kohn,~W.; Sham,~L.~J. Self-consistent equations including exchange and correlation effects. \emph{Physical review} \textbf{1965}, \emph{140}, A1133\relax
\mciteBstWouldAddEndPuncttrue
\mciteSetBstMidEndSepPunct{\mcitedefaultmidpunct}
{\mcitedefaultendpunct}{\mcitedefaultseppunct}\relax
\EndOfBibitem
\bibitem[Parr(1980)]{parr1980density}
Parr,~R.~G. Density functional theory of atoms and molecules. Horizons of Quantum Chemistry: Proceedings of the Third International Congress of Quantum Chemistry Held at Kyoto, Japan, October 29-November 3, 1979. 1980; pp 5--15\relax
\mciteBstWouldAddEndPuncttrue
\mciteSetBstMidEndSepPunct{\mcitedefaultmidpunct}
{\mcitedefaultendpunct}{\mcitedefaultseppunct}\relax
\EndOfBibitem
\bibitem[Huron \latin{et~al.}(1973)Huron, Malrieu, and Rancurel]{huron1973iterative}
Huron,~B.; Malrieu,~J.; Rancurel,~P. Iterative perturbation calculations of ground and excited state energies from multiconfigurational zeroth-order wavefunctions. \emph{The Journal of Chemical Physics} \textbf{1973}, \emph{58}, 5745--5759\relax
\mciteBstWouldAddEndPuncttrue
\mciteSetBstMidEndSepPunct{\mcitedefaultmidpunct}
{\mcitedefaultendpunct}{\mcitedefaultseppunct}\relax
\EndOfBibitem
\bibitem[Sharma \latin{et~al.}(2017)Sharma, Holmes, Jeanmairet, Alavi, and Umrigar]{sharma2017semistochastic}
Sharma,~S.; Holmes,~A.~A.; Jeanmairet,~G.; Alavi,~A.; Umrigar,~C.~J. Semistochastic heat-bath configuration interaction method: Selected configuration interaction with semistochastic perturbation theory. \emph{Journal of chemical theory and computation} \textbf{2017}, \emph{13}, 1595--1604\relax
\mciteBstWouldAddEndPuncttrue
\mciteSetBstMidEndSepPunct{\mcitedefaultmidpunct}
{\mcitedefaultendpunct}{\mcitedefaultseppunct}\relax
\EndOfBibitem
\bibitem[Garniron \latin{et~al.}(2018)Garniron, Scemama, Giner, Caffarel, and Loos]{garniron2018selected}
Garniron,~Y.; Scemama,~A.; Giner,~E.; Caffarel,~M.; Loos,~P.-F. Selected configuration interaction dressed by perturbation. \emph{The Journal of Chemical Physics} \textbf{2018}, \emph{149}\relax
\mciteBstWouldAddEndPuncttrue
\mciteSetBstMidEndSepPunct{\mcitedefaultmidpunct}
{\mcitedefaultendpunct}{\mcitedefaultseppunct}\relax
\EndOfBibitem
\bibitem[Martin(2003)]{martin2003natural}
Martin,~R.~L. Natural transition orbitals. \emph{The Journal of chemical physics} \textbf{2003}, \emph{118}, 4775--4777\relax
\mciteBstWouldAddEndPuncttrue
\mciteSetBstMidEndSepPunct{\mcitedefaultmidpunct}
{\mcitedefaultendpunct}{\mcitedefaultseppunct}\relax
\EndOfBibitem
\bibitem[Sun(2015)]{sun2015libcint}
Sun,~Q. Libcint: An efficient general integral library for gaussian basis functions. \emph{Journal of computational chemistry} \textbf{2015}, \emph{36}, 1664--1671\relax
\mciteBstWouldAddEndPuncttrue
\mciteSetBstMidEndSepPunct{\mcitedefaultmidpunct}
{\mcitedefaultendpunct}{\mcitedefaultseppunct}\relax
\EndOfBibitem
\bibitem[Sun \latin{et~al.}(2018)Sun, Berkelbach, Blunt, Booth, Guo, Li, Liu, McClain, Sayfutyarova, Sharma, \latin{et~al.} others]{sun2018pyscf}
Sun,~Q.; Berkelbach,~T.~C.; Blunt,~N.~S.; Booth,~G.~H.; Guo,~S.; Li,~Z.; Liu,~J.; McClain,~J.~D.; Sayfutyarova,~E.~R.; Sharma,~S.; others PySCF: the Python-based simulations of chemistry framework. \emph{Wiley Interdisciplinary Reviews: Computational Molecular Science} \textbf{2018}, \emph{8}, e1340\relax
\mciteBstWouldAddEndPuncttrue
\mciteSetBstMidEndSepPunct{\mcitedefaultmidpunct}
{\mcitedefaultendpunct}{\mcitedefaultseppunct}\relax
\EndOfBibitem
\bibitem[Sun \latin{et~al.}(2020)Sun, Zhang, Banerjee, Bao, Barbry, Blunt, Bogdanov, Booth, Chen, Cui, \latin{et~al.} others]{sun2020recent}
Sun,~Q.; Zhang,~X.; Banerjee,~S.; Bao,~P.; Barbry,~M.; Blunt,~N.~S.; Bogdanov,~N.~A.; Booth,~G.~H.; Chen,~J.; Cui,~Z.-H.; others Recent developments in the PySCF program package. \emph{The Journal of chemical physics} \textbf{2020}, \emph{153}, 024109\relax
\mciteBstWouldAddEndPuncttrue
\mciteSetBstMidEndSepPunct{\mcitedefaultmidpunct}
{\mcitedefaultendpunct}{\mcitedefaultseppunct}\relax
\EndOfBibitem
\bibitem[Piecuch \latin{et~al.}(2006)Piecuch, W{\l}och, Gour, and Kinal]{piecuch2006single}
Piecuch,~P.; W{\l}och,~M.; Gour,~J.~R.; Kinal,~A. Single-reference, size-extensive, non-iterative coupled-cluster approaches to bond breaking and biradicals. \emph{Chemical physics letters} \textbf{2006}, \emph{418}, 467--474\relax
\mciteBstWouldAddEndPuncttrue
\mciteSetBstMidEndSepPunct{\mcitedefaultmidpunct}
{\mcitedefaultendpunct}{\mcitedefaultseppunct}\relax
\EndOfBibitem
\bibitem[Piecuch and W{\l}och(2005)Piecuch, and W{\l}och]{piecuch2005renormalized}
Piecuch,~P.; W{\l}och,~M. Renormalized coupled-cluster methods exploiting left eigenstates of the similarity-transformed Hamiltonian. \emph{The Journal of chemical physics} \textbf{2005}, \emph{123}, 224105\relax
\mciteBstWouldAddEndPuncttrue
\mciteSetBstMidEndSepPunct{\mcitedefaultmidpunct}
{\mcitedefaultendpunct}{\mcitedefaultseppunct}\relax
\EndOfBibitem
\bibitem[Piecuch \latin{et~al.}(2015)Piecuch, Hansen, and Ajala]{piecuch2015benchmarking}
Piecuch,~P.; Hansen,~J.~A.; Ajala,~A.~O. Benchmarking the completely renormalised equation-of-motion coupled-cluster approaches for vertical excitation energies. \emph{Molecular Physics} \textbf{2015}, \emph{113}, 3085--3127\relax
\mciteBstWouldAddEndPuncttrue
\mciteSetBstMidEndSepPunct{\mcitedefaultmidpunct}
{\mcitedefaultendpunct}{\mcitedefaultseppunct}\relax
\EndOfBibitem
\bibitem[Piecuch \latin{et~al.}(2002)Piecuch, Kucharski, Kowalski, and Musia{\l}]{piecuch2002efficient}
Piecuch,~P.; Kucharski,~S.~A.; Kowalski,~K.; Musia{\l},~M. Efficient computer implementation of the renormalized coupled-cluster methods: the r-ccsd [t], r-ccsd (t), cr-ccsd [t], and cr-ccsd (t) approaches. \emph{Computer Physics Communications} \textbf{2002}, \emph{149}, 71--96\relax
\mciteBstWouldAddEndPuncttrue
\mciteSetBstMidEndSepPunct{\mcitedefaultmidpunct}
{\mcitedefaultendpunct}{\mcitedefaultseppunct}\relax
\EndOfBibitem
\bibitem[Kowalski and Piecuch(2004)Kowalski, and Piecuch]{kowalski2004new}
Kowalski,~K.; Piecuch,~P. New coupled-cluster methods with singles, doubles, and noniterative triples for high accuracy calculations of excited electronic states. \emph{The Journal of Chemical Physics} \textbf{2004}, \emph{120}, 1715--1738\relax
\mciteBstWouldAddEndPuncttrue
\mciteSetBstMidEndSepPunct{\mcitedefaultmidpunct}
{\mcitedefaultendpunct}{\mcitedefaultseppunct}\relax
\EndOfBibitem
\bibitem[Piecuch \latin{et~al.}(2009)Piecuch, Gour, and W{\l}och]{piecuch2009left}
Piecuch,~P.; Gour,~J.~R.; W{\l}och,~M. Left-eigenstate completely renormalized equation-of-motion coupled-cluster methods: Review of key concepts, extension to excited states of open-shell systems, and comparison with electron-attached and ionized approaches. \emph{International Journal of Quantum Chemistry} \textbf{2009}, \emph{109}, 3268--3304\relax
\mciteBstWouldAddEndPuncttrue
\mciteSetBstMidEndSepPunct{\mcitedefaultmidpunct}
{\mcitedefaultendpunct}{\mcitedefaultseppunct}\relax
\EndOfBibitem
\bibitem[Fradelos \latin{et~al.}(2011)Fradelos, Lutz, Weso{\l}owski, Piecuch, and W{\l}och]{fradelos2011embedding}
Fradelos,~G.; Lutz,~J.~J.; Weso{\l}owski,~T.~A.; Piecuch,~P.; W{\l}och,~M. Embedding vs supermolecular strategies in evaluating the hydrogen-bonding-induced shifts of excitation energies. \emph{Journal of Chemical Theory and Computation} \textbf{2011}, \emph{7}, 1647--1666\relax
\mciteBstWouldAddEndPuncttrue
\mciteSetBstMidEndSepPunct{\mcitedefaultmidpunct}
{\mcitedefaultendpunct}{\mcitedefaultseppunct}\relax
\EndOfBibitem
\bibitem[Schmidt \latin{et~al.}(1993)Schmidt, Baldridge, Boatz, Elbert, Gordon, Jensen, Koseki, Matsunaga, Nguyen, Su, \latin{et~al.} others]{schmidt1993general}
Schmidt,~M.~W.; Baldridge,~K.~K.; Boatz,~J.~A.; Elbert,~S.~T.; Gordon,~M.~S.; Jensen,~J.~H.; Koseki,~S.; Matsunaga,~N.; Nguyen,~K.~A.; Su,~S.; others General atomic and molecular electronic structure system. \emph{Journal of computational chemistry} \textbf{1993}, \emph{14}, 1347--1363\relax
\mciteBstWouldAddEndPuncttrue
\mciteSetBstMidEndSepPunct{\mcitedefaultmidpunct}
{\mcitedefaultendpunct}{\mcitedefaultseppunct}\relax
\EndOfBibitem
\bibitem[Barca \latin{et~al.}(2020)Barca, Bertoni, Carrington, Datta, De~Silva, Deustua, Fedorov, Gour, Gunina, Guidez, \latin{et~al.} others]{barca2020recent}
Barca,~G.~M.; Bertoni,~C.; Carrington,~L.; Datta,~D.; De~Silva,~N.; Deustua,~J.~E.; Fedorov,~D.~G.; Gour,~J.~R.; Gunina,~A.~O.; Guidez,~E.; others Recent developments in the general atomic and molecular electronic structure system. \emph{The Journal of chemical physics} \textbf{2020}, \emph{152}, 154102\relax
\mciteBstWouldAddEndPuncttrue
\mciteSetBstMidEndSepPunct{\mcitedefaultmidpunct}
{\mcitedefaultendpunct}{\mcitedefaultseppunct}\relax
\EndOfBibitem
\bibitem[Johnson~III(2022)]{johnson2022nist}
Johnson~III,~R. NIST computational chemistry comparison and benchmark database. May 2022 ed. \emph{NIST Standard Reference Database} \textbf{2022}, \relax
\mciteBstWouldAddEndPunctfalse
\mciteSetBstMidEndSepPunct{\mcitedefaultmidpunct}
{}{\mcitedefaultseppunct}\relax
\EndOfBibitem
\bibitem[Loos \latin{et~al.}(2018)Loos, Scemama, Blondel, Garniron, Caffarel, and Jacquemin]{loos2018mountaineering}
Loos,~P.-F.; Scemama,~A.; Blondel,~A.; Garniron,~Y.; Caffarel,~M.; Jacquemin,~D. A mountaineering strategy to excited states: Highly accurate reference energies and benchmarks. \emph{Journal of chemical theory and computation} \textbf{2018}, \emph{14}, 4360--4379\relax
\mciteBstWouldAddEndPuncttrue
\mciteSetBstMidEndSepPunct{\mcitedefaultmidpunct}
{\mcitedefaultendpunct}{\mcitedefaultseppunct}\relax
\EndOfBibitem
\bibitem[Thouless(1960)]{thouless1960stability}
Thouless,~D.~J. Stability conditions and nuclear rotations in the Hartree-Fock theory. \emph{Nuclear Physics} \textbf{1960}, \emph{21}, 225--232\relax
\mciteBstWouldAddEndPuncttrue
\mciteSetBstMidEndSepPunct{\mcitedefaultmidpunct}
{\mcitedefaultendpunct}{\mcitedefaultseppunct}\relax
\EndOfBibitem
\bibitem[Schulz(1973)]{schulz1973resonances}
Schulz,~G.~J. Resonances in electron impact on diatomic molecules. \emph{Reviews of Modern Physics} \textbf{1973}, \emph{45}, 423\relax
\mciteBstWouldAddEndPuncttrue
\mciteSetBstMidEndSepPunct{\mcitedefaultmidpunct}
{\mcitedefaultendpunct}{\mcitedefaultseppunct}\relax
\EndOfBibitem
\bibitem[Balkov{\'a} and Bartlett(1992)Balkov{\'a}, and Bartlett]{balkova1992coupled}
Balkov{\'a},~A.; Bartlett,~R.~J. Coupled-cluster method for open-shell singlet states. \emph{Chemical physics letters} \textbf{1992}, \emph{193}, 364--372\relax
\mciteBstWouldAddEndPuncttrue
\mciteSetBstMidEndSepPunct{\mcitedefaultmidpunct}
{\mcitedefaultendpunct}{\mcitedefaultseppunct}\relax
\EndOfBibitem
\bibitem[Lutz \latin{et~al.}(2018)Lutz, Nooijen, Perera, and Bartlett]{lutz2018reference}
Lutz,~J.~J.; Nooijen,~M.; Perera,~A.; Bartlett,~R.~J. Reference dependence of the two-determinant coupled-cluster method for triplet and open-shell singlet states of biradical molecules. \emph{The Journal of chemical physics} \textbf{2018}, \emph{148}, 164102\relax
\mciteBstWouldAddEndPuncttrue
\mciteSetBstMidEndSepPunct{\mcitedefaultmidpunct}
{\mcitedefaultendpunct}{\mcitedefaultseppunct}\relax
\EndOfBibitem
\end{mcitethebibliography}

\clearpage
\onecolumngrid

\section{Supporting Information}
\renewcommand{\thesection}{S\arabic{section}}
\renewcommand{\theequation}{S\arabic{equation}}
\renewcommand{\thefigure}{S\arabic{figure}}
\renewcommand{\thetable}{S\arabic{table}}
\setcounter{section}{0}
\setcounter{figure}{0}
\setcounter{equation}{0}
\setcounter{table}{0}
\section{Charge Transfer Excitation Energies}
\setlength{\tabcolsep}{5pt}

\begin{table}[h!]
    \centering
    \begin{tabular}{l l S[table-format=2.2] S[table-format=2.2] S[table-format=2.2] S[table-format=2.2]}
         Molecule    & State & {ESMP2} & {ASCC} & {EOM-CCSD} & {$\delta$-CR-EOM-CC(2,3),A} \\ \hline
                     lithium fluoride & h1p1 &  6.60 &  6.39 &  6.36 &  6.36 \\                                         
                                      & h2p1 &  7.01 &  6.84 &  6.82 &  6.82 \\                                         
                         chloroethylene & h1p1 &  7.01 &  7.31 &  7.22 &  6.98 \\                                         
                                      & h2p1 &  7.51 &  7.81 &  7.92 &  7.67 \\
                            acrolein  & h1p1 &  3.78 &  3.81 &  3.92 &  3.62 \\                                         
                                      & h2p1 &  6.16 &  6.86 &  6.88 &  6.49 \\                                         
      ammonia $\rightarrow$ difluorine & h1p1 &  7.62 &  8.21 &  8.82 &  8.18 \\                                         
     chloride $\rightarrow$ dinitrogen & h1p2 &  5.88 &  5.97 &  6.18 &  5.90 \\                                         
                                      & h1p1 &  5.68 &  5.87 &  6.19 &  5.86 \\                                         
                                      & h3p1 &  5.78 &  5.97 &  6.29 &  5.98 \\                                         
                                      & h2p2 &  5.93 &  6.05 &  6.30 &  5.98 \\                                         
                                      & h2p1 &  5.72 &  5.98 &  6.30 &  5.98 \\                                         
                                      & h3p2 &  5.91 &  6.03 &  6.30 &  5.98 \\                                         
chloride $\rightarrow$ carbon monoxide & h1p1 &  5.16 &  5.20 &  5.39 &  5.18 \\                                         
                                      & h1p2 &  5.22 &  5.27 &  5.44 &  5.23 \\                                         
                                      & h3p1 &  5.24 &  5.27 &  5.46 &  5.26 \\                                         
                                      & h2p1 &  5.19 &  5.27 &  5.47 &  5.27 \\                                         
                                      & h2p2 &  5.32 &  5.34 &  5.52 &  5.32 \\                                         
                                      & h3p2 &  5.30 &  5.35 &  5.53 &  5.32 \\                                         
         chloride $\rightarrow$ ethylene & h1p1 &  5.08 &  5.33 &  5.55 &  5.33 \\                                         
                                      & h2p1 &  5.25 &  5.48 &  5.69 &  5.44 \\                                         
                                      & h3p1 &  5.16 &  5.48 &  5.71 &  5.46 \\                                         
    \end{tabular}
    \caption{Vertical excitation energies in eV for charge transfer systems}
    \label{tab: sict}
\end{table}

\clearpage
\section{Molecular Geometries}

The supporting information for the QUEST benchmark contains geometries for the molecules from the QUEST set. The remaining charge transfer geometries are reported below in angstroms.

\noindent \bf{lithium fluoride}
\begin{verbatim}
F   0.0000000    0.0000000    0.3968040                                                      
Li  0.0000000    0.0000000   -1.1904110
\end{verbatim}

\noindent \bf{chloroethylene}
\begin{verbatim}
C   0.0000000    0.7588320    0.0000000                                                      
C   1.2921560    1.0327720    0.0000000                                                      
Cl -0.6259060   -0.8573720    0.0000000                                                      
H  -0.7694050    1.5061990    0.0000000                                                      
H   2.0399880    0.2605710    0.0000000                                                      
H   1.6168930    2.0589350    0.0000000
\end{verbatim}

\noindent \bf{acrolein}
\begin{verbatim}
C  -0.1533010   -0.7559140    0.0000000                                                     
C   0.0000000    0.7250820    0.0000000                                                     
C   1.2220650    1.2941010    0.0000000                                                     
O  -1.2237110   -1.3289000    0.0000000                                                     
H   0.8134290   -1.3220750    0.0000000                                                     
H  -0.9232470    1.3153200    0.0000000                                                     
H   1.3568600    2.3802520    0.0000000                                                     
H   2.1300570    0.6780920    0.0000000
\end{verbatim}

\noindent \bf{ammonia $\rightarrow$ difluorine}
\begin{verbatim}
N   0.0000000    0.0000000    0.1277920                                                      
H   0.0000000    0.9318900   -0.2981820                                                     
H   0.8070400   -0.4659450   -0.2981820                                                     
H  -0.8070400   -0.4659450   -0.2981820                                                     
F   0.0000000    0.0000000    6.1277920                                                      
F   0.0000000    0.0000000    7.5597920
\end{verbatim}

\noindent \bf{chloride $\rightarrow$ dinitrogen}
\begin{verbatim}
N   0.0000000    0.0000000   -0.5560000                                                      
N   0.0000000    0.0000000    0.5560000                                                      
Cl  0.0000000    4.0000000    0.0000000 
\end{verbatim}

\noindent \bf{chloride $\rightarrow$ carbon monoxide}
\begin{verbatim}
C   0.0000000    0.0000000   -0.5690000                                                     
O   0.0000000    0.0000000    0.5690000                                                     
Cl  0.0000000    4.0000000    0.0000000 
\end{verbatim}

\noindent \bf{chloride $\rightarrow$ ethylene}
\begin{verbatim}
C   0.0000000    0.0000000    0.6726230
C   0.0000000    0.0000000   -0.6726230
H   0.9341310    0.0000000    1.2459530
H  -0.9341310    0.0000000    1.2459530
H  -0.9341310    0.0000000   -1.2459530
H   0.9341310    0.0000000   -1.2459530
Cl  0.0000000    4.0000000    0.0000000
\end{verbatim}

\renewcommand{\thesection}{\Roman{section}}
\setcounter{section}{6}

\end{document}